\documentclass[12pt, preprint]{aastex}

\begin{document}

\title{Three-dimensional Radiative Transfer Modeling of the Polarization of the Sun's Continuous Spectrum} 

\shorttitle{Modeling the Polarization of the Sun's Continuous Spectrum}

\author{Javier Trujillo Bueno\altaffilmark{1,2} and Nataliya Shchukina\altaffilmark{3}}
\altaffiltext{1}{Instituto de Astrof\'{\i}sica de Canarias, 38205 La Laguna,
Tenerife, Spain}
\altaffiltext{2}{Consejo Superior de Investigaciones Cient\'{\i}ficas (Spain)}
\altaffiltext{3}{Main Astronomical Observatory, National Academy of Sciences, 27 Zabolotnogo Street, Kiev 03680, Ukraine}\email{jtb@iac.es, shchukin@mao.kiev.ua}

\date{{\bf Accepted in December 2008 for publication in The Astrophysical Journal}}

\begin{abstract}

Polarized light provides the most reliable source of information at our disposal 
for diagnosing the physical properties of astrophysical plasmas, including the three-dimensional (3D) structure of the solar atmosphere. Here we formulate and solve the 3D radiative transfer problem of the
linear polarization of the solar continuous radiation, which is principally produced by Rayleigh and Thomson scattering. Our approach takes into account not only the anisotropy of the solar continuum radiation, but also the symmetry-breaking effects caused by the horizontal atmospheric inhomogeneities produced by the solar surface convection. We show that such symmetry-breaking effects do produce observable signatures in $Q/I$ and $U/I$, even at the very center of the solar disk where we observe the forward scattering case, but their detection would require obtaining very-high-resolution linear polarization images of the solar surface. Without spatial and/or temporal resolution $U/I{\approx}0$ and the only observable quantity is $Q/I$, whose wavelength variation at a solar disk position close to the limb has been recently determined semi-empirically. Interestingly, our 3D radiative transfer modeling of the polarization of the Sun's continuous spectrum in a well-known 3D hydrodynamical model of the solar photosphere shows remarkable agreement with the semi-empirical determination, significantly better than that obtained via the use of one-dimensional (1D) atmospheric models. Although this result confirms that the above-mentioned 3D model was indeed a suitable choice for our Hanle-effect estimation of the substantial amount of ``hidden" magnetic energy that is stored in the quiet solar photosphere, we have found however some small discrepancies whose origin may be due to uncertainties in the semi-empirical data and/or in the thermal and density structure of the 3D model. For this reason, we have paid some attention also to other (more familiar) observables, like the center-limb variation of the continuum intensity, which we have calculated taking into account the scattering contribution to the continuum source function. The overall agreement with the observed center-limb variation turns out to be impressive, but we find a hint that the model's temperature gradients in the continuum forming layers could be {\em slightly} too steep, perhaps because all current simulations of solar surface convection and magnetoconvection compute the radiative flux divergence ignoring the fact that the effective polarizability is not completely negligible, especially in the downward-moving intergranular lane plasma.

\end{abstract}


\section{Introduction}

The main aim of this investigation is to formulate and solve the 3D radiative transfer problem of the linear polarization of the Sun's continuous spectrum, which results principally from Rayleigh scattering at neutral hydrogen and Thomson scattering at free electrons. Our scientific motivation is twofold. On the one hand, we want to use the polarization of the Sun's continuous radiation as a diagnostic tool of the thermal and density structure of the solar photosphere. On the other hand, a realistic modeling of the continuum polarization is important for quantitative analysis of the linearly-polarized solar limb spectrum. 

As we shall see below, the polarization of the solar continuum radiation is largely determined by the effective polarizability and by the radiation field's anisotropy, which in turn depend on the density and thermal structure of the stellar atmosphere under consideration. Collisional and/or magnetic depolarization do not play any role on the polarization of the continuum radiation of the Sun's visible spectrum. Therefore, the more realistic is the thermal and density structure of a given solar atmospheric model, the closer to the empirical data will be the calculated linear polarization of the solar continuum radiation.   

While our present telescopes and polarimeters can reach very high accuracy in determining the relative polarization variations between line and continuum features in the solar spectrum (e.g., Stenflo \& Keller 1997), they are not suitable for  determining the zero-point of the polarization scale with a similar polarization precision. Nevertheless, Stenflo (2005) could extract the polarization of the Sun's continuous spectrum from Gandorfer's (2000, 2002, 2005) atlas of the linearly-polarized solar limb spectrum. As emphasized by Ivanov (1991), this so-called second solar spectrum contains multitude of fractional linear polarization ($Q/I$) features seen both ``in emission" and ``in absorption", i.e., with correspondingly larger and smaller polarization than in the adjacent continuum\footnote{We recall that the Stokes $I$ parameter represents the intensity at a given wavelength, while Stokes $Q$ is the intensity difference between linear polarization parallel and perpendicular to a given reference direction (e.g., the direction perpendicular to the straight line joining the disk center with the observed point). Stokes $U$ would be then the intensity difference between linear polarization at $+45^{\circ}$ and $-45^{\circ}$ with respect to the chosen reference direction, where positive angles are measured counterclockwise by an observer facing the radiation.}. While many of such $Q/I$ signals are dominated by the {\em selective emission} and {\em selective absorption} processes caused by the radiatively induced quantum coherences in the energy levels of the atoms and molecules of the solar atmosphere (e.g., Trujillo Bueno 2001), many of the ``absorption" features result from depolarizing line contributions which depress the continuum polarization level. Stenflo's (2005) determination of the polarization of the Sun's continuous spectrum is {\em semi-empirical} because it was based on a model for the behavior of these spectral lines that depolarize the continuum polarization. His strategy consisted in choosing spectral windows dominated by purely depolarizing lines and in modeling the observed relative line depth in $Q/I$ in terms of the relative line depth in Stokes $I$. He pointed out that his semi-empirically determined continuum polarization lies systematically lower than the values predicted by Fluri \& Stenflo (1999) for $\lambda{>}4000$ \AA\ from radiative transfer modeling in 1D  semi-empirical models of the solar atmosphere. 

Probably, the most interesting result of Stenflo's (2005) semi-empirical determination of the wavelength dependence of the continuum polarization at $\mu={\rm cos}{\theta}\,{\approx}\,0.1$ (with $\theta$ the heliocentric angle) is the confirmation of the 1979 theoretical expectation by L. Auer (see Stenflo, Twerenbold \& Harvey 1983; Stenflo 2003) that it must show a prominent jump around the Balmer limit at 3646 \AA. At larger  wavelengths the dominant contribution to the polarization of the solar continuous spectrum is Rayleigh scattering at neutral hydrogen (Chandrasekhar 1960), which in turn is dominated by scattering in the distant line wings of the Lyman series lines (Stenflo 2005). At visible wavelengths this Rayleigh scattering at neutral hydrogen is the second most important opacity source after that of the (non-polarizing) H$^{-}$ absorption. As it happens with other observables of the Sun's continuous spectrum (e.g., the center-limb variation and the contrast of solar granulation) a prominent jump should be seen also in the scattering polarization when the extra contribution to the continuum opacity due to Balmer bound-free transitions becomes operative, because this contribution is larger than the Rayleigh scattering one. Interestingly enough, Stenflo's (2005) determination shows that this Balmer jump is indeed produced in the polarization of the Sun's continuous spectrum, but approximately at a wavelength 80 \AA\ larger than the nominal position of the Balmer series limit. According to Stenflo (2005) this can be understood in terms of Stark broadening of the Balmer bound-bound transitions, which become very crowded when approaching the Balmer limit and merge into a quasi-continuum whose ensuing opacity becomes larger than the Rayleigh scattering opacity well before the Balmer series limit is reached. 

In this paper we take the opportunity of contrasting Stenflo's (2005) determination of the polarization of the Sun's continuous spectrum between 3000 \AA\ and 7000 \AA\ with detailed radiative transfer simulations in a ``realistic"  3D snapshot model of the solar photosphere, which resulted from Asplund et al.'s (2000) hydrodynamical simulations of solar surface convection. This is the same 3D snapshot model we used in our investigations of the Hanle effect in the Sr {\sc i} line at 4607 \AA, which allowed us to conclude that in the ``quiet" regions of the solar photosphere there is a substantial amount of hidden magnetic energy carried by tangled magnetic fields at subresolution scales (Trujillo Bueno, Shchukina \& Asensio Ramos 2004; Trujillo Bueno, Asensio Ramos \& Shchukina 2006; Trujillo Bueno \& Shchukina 2007). As we shall see below, our 3D radiative transfer modeling of the polarization of the Sun's continuous spectrum is in very good agreement with Stenflo's (2005) semi-empirical determination. However, there are some small discrepancies, whose origin may be due to various possible reasons (e.g., uncertainties in Stenflo's  semi-empirical determination and/or in the thermal and density structure of the 3D model). In order to investigate this issue from a complementary perspective, we have considered also other (more familiar) observables, like the center-limb variation of the continuum intensity and histograms of the intensity fluctuations produced by the solar granulation pattern, which we have calculated taking into account the scattering contribution to the continuum source function. 

The outline of this paper is the following. The formulation of the continuum polarization problem in a 3D medium like that of the solar atmosphere is presented in \S2, where we establish the relevant equations and mention how we have solved them. After illustrating and discussing in \S3 the atmospheric models we have chosen for this investigation, we show in \S4 the spatial variation of the radiation field tensors (which quantify the symmetry properties of the radiation field), while in \S5 we show both the spatial and the wavelength variation of the effective polarizability. The center-limb variation of the Sun's continuous spectrum is carefully considered in \S6, while \S7 focuses on confronting the calculated histogram of the intensity distribution of the model's granulation with that observed on the Sun with the {\em Hinode} space telescope. \S8 presents a careful investigation of the wavelength dependence of the anisotropy factor, which is one of the fundamental quantities for scattering polarization. Our calculations of the polarization of the Sun's continuous spectrum are presented in \S9, where we compare them with Stenflo's (2005) semi-empirical determination. The interesting polarization effects produced by the breaking of the axial symmetry of the photospheric radiation field are discussed in \S10. Finally, \S11 summarizes our main conclusions with an outlook to future research.

\section{The physical mechanisms and the relevant equations}

As it happens with the (polarizing) spectral lines of the second solar spectrum, there are 
two mechanisms capable of introducing linear polarization in the continuous spectrum of a stellar atmosphere: scattering and dichroism. Scattering processes produce a {\em selective emission} of polarization components, while dichroism causes a {\em selective absorption} of polarization components. 

In the solar atmosphere the scattering processes that make the dominant contribution to the polarization of the continuous spectrum between 3000 \AA\ and 7000 \AA\ are scattering at neutral hydrogen in its ground state (Lyman scattering), and to a smaller degree Thomson scattering at free electrons (e.g., Chandrasekhar 1960; Landi Degl'Innocenti \& Landolfi 2004; Stenflo 2005). Dichroism requires that the (degenerate) bound level from which a photoionization or scattering process takes place is polarized (i.e., its sublevels must be unequally populated and/or harbor quantum interferences between them), something that in the ``quiet" regions of the solar atmosphere can occur through anisotropic radiative pumping processes (e.g., Trujillo Bueno 2001). At wavelengths larger than that of  the Balmer limit at 3646 \AA\ the contribution of dichroism to the polarization of the continuous spectrum is expected to be negligible, because in the visible and near-IR regions the major contribution to the continuum absorption coefficient comes from the photoionization of the H$^{-}$ ion, whose single bound level cannot be polarized because its total angular momentum is zero. However, at smaller wavelengths there is a significant opacity contribution due to bound-free transitions from the lower level of the Balmer lines and from bound levels of other chemical elements like C, Si, Fe, Mg and Al, so that dichroism could in principle have an impact on the linear polarization of the emergent continuum radiation, given that such bound levels can in principle be polarized (cf. Landi Degl'Innocenti \& Landolfi 2004). The investigation of this possibility would require solving the so-called non-LTE problem of the $2nd$ kind in the 3D photospheric model used here, using a realistic multilevel model for each of such species. Although such type of calculations could be performed with the radiative transfer computer program of Manso Sainz \& Trujillo Bueno (2003), we leave such a challenging investigation for a future occasion. In any case, for the wavelength range considered in this paper the influence of dichroism on the continuum polarization is probably totally negligible. This is because of the three lower levels of the Balmer lines the most populated one (i.e., the metastable level ${\rm S}_{1/2}$) cannot be aligned, and the atomic polarization of the only lower level of the Balmer lines that can be aligned (i.e., the state ${\rm P}_{3/2}$) is expected to be insignificant at the relatively low photospheric heights where the monochromatic optical depth is unity along the line of sight (LOS).  

In summary, in this paper we assume that the polarization of the Sun's continuous radiation is caused by Thomson scattering at free electrons and by Rayleigh scattering from the ground level of neutral hydrogen, both of which have polarizability unity (e.g., Chandrasekhar 1960; Stenflo 2005). The contribution of these processes to the total absorption coefficient ($\chi_c$) is quantified by

\begin{equation}
{\sigma}_c\,=\,{\sigma}_T\,N_e+\,{\sigma}_R\,n_1(H),
\end{equation}
where $\sigma_T=6.653{\times}10^{-25}\,{\rm cm}^2$ is the Thomson scattering cross section (which is independent of wavelength), $N_e$ the electron number density, $n_1(H)$ the population of the ground level of hydrogen (which in the solar photosphere is practically identical to the total neutral hydrogen number density), 
and $\sigma_R$ the wavelength-dependent Rayleigh cross section. 
For $\lambda{>}2000$ \AA\ a very good approximation for $\sigma_R$ is (Baschek \& Scholz 1982)

\begin{equation}
{\sigma}_R\,{\approx}\,{\sigma_T}{(}{{966}\over{\lambda}}{)}^4{[}1+{(}{{1566}\over{\lambda}}{)}^2+{(}{{1480}\over{\lambda}}{)}^4{]},
\end{equation} 
with $\lambda$ in \AA. The total absorption coefficient is given by

\begin{equation}
{\chi}_c\,=\kappa_c\,+\,{\sigma}_c,
\end{equation}
where $\kappa_c$ contains all the relevant non-scattering contributions to the continuum absorption coefficient. At visible and near-IR wavelengths the main contribution to $\kappa_c$ is caused by bound-free transitions in the H$^{-}$ ion, whose single bound level is the $1{s^2}\,{^2}{\rm S}_0$. We have included also the contributions due to free-free transitions in H$^{-}$, 
bound-free and free-free transitions in hydrogen, and those due to bound-free transitions in carbon, silicon, iron, magnesium and aluminium with the populations of the bound levels (including that of the H$^{-}$ ion) calculated assuming LTE, which is a fairly good approximation in the lower solar photosphere (e.g., Vernazza, Avrett \& Loeser 1981). However, in this 3D radiative transfer investigation we have not included the Balmer bound-bound transitions, which would be needed for modeling the $80$\AA\ shift of the Balmer jump towards larger wavelengths that we mentioned in \S1.  

Since in the wavelength range considered here the contribution of dichroism is expected to be negligible, the absorption coefficient does not depend on the polarization of the incident radiation and the transfer equation for the Stokes parameter $X$ (with $X=I,Q,U$) at a given frequency $\nu$ and direction of propagation $\vec{\Omega}$ is given by 

\begin{equation}
{{d}\over{d\tau}}X\,=\,X\,-\,S_X,
\end{equation}
where $\tau$ (with ${d\tau}=-\chi_c\,ds$) is the monochromatic optical distance along the ray. Note also that an approximate expression for estimating the emergent fractional linear polarization is $X/I{\approx}S_X/{S}_I$, with the corresponding source function values calculated at the atmospheric height where the optical depth is unity along the LOS.

The expressions for $S_X$ (with $X=I,Q,U$) can be easily obtained as a particular case of those derived by Trujillo Bueno \& Manso Sainz (1999) and 
Manso Sainz \& Trujillo Bueno (1999), which correspond to the case of scattering polarization in a resonance line without lower-level polarization. Such equations were applied by Trujillo Bueno \& Shchukina (2007) for investigating the 3D scattering polarization problem in the Sr {\sc i} 4607 \AA\ line. In order to obtain the $S_X$ expressions for the continuum polarization problem in a 3D medium, it suffices with choosing the following particular values for the indicated quantities: $w^{(2)}_{J_uJ_l}=1$ (because the polarizability of Lyman and Thomson scattering is unity), ${\cal H}^{(2)}=1$ (because in the absence of dichroism there is no influence of the Hanle effect on the continuum polarization), ${\phi}(x)=1$ (because we are considering continuum radiation only, so that there is no frequency integration over a line profile, given that each continuum frequency is treated independently), ${\delta}^{(2)}=0$ (because in the absence of dichroism there is no influence of depolarizing collisions on the polarization of the continuum radiation), $\epsilon=0$ (because for  $\kappa_c=0$ the resulting equations must be those of a purely scattering atmosphere), $\chi_l=\sigma_c$ (because we are not considering line polarization, but continuum polarization caused by Rayleigh and Thomson scattering) and $\chi_c=\kappa_c$ (simply because we are using here a different notation for the non-polarizing part of the continuum absorption coefficient). With these particularizations it is straightforward to find that\footnote{We point out that these expressions can be easily written in a more compact notation by using the ${\cal{T}}^K_Q(i,{\bf {\Omega}})$ tensor defined in Table 5.6 of Landi Degl'Innocenti \& Landolfi (2004).}

\begin{eqnarray}
  {S}_I={{\kappa_c}\over{\kappa_c+\sigma_c}}B_{\nu}(T)\,+\,
  {{\sigma_c}\over{\kappa_c+\sigma_c}}\,\Big{\{}
  {{J}^0_0}+\frac{1}{2\sqrt{2}}(3\mu^2-1){{J}^2_0} \nonumber 
  \end{eqnarray}
\vspace{-0.1in}
\begin{eqnarray}
  - \sqrt{3}  \mu \sqrt{1-\mu^2} (\cos \chi
  {{{\tilde J}}^2_1} + \sin 
  \chi {{{\hat J}}^2_1}) \nonumber 
\end{eqnarray}
\vspace{-0.2in}
\begin{eqnarray}
\hspace{1.0in} +\frac{\sqrt{3}}{2} (1-\mu^2) (\cos
  2\chi \, {{{\tilde J}}^2_2}+\sin 2\chi \, 
  {{{\hat J}}^2_2}) \Big{\}},
  \end{eqnarray} 

\begin{eqnarray}
  {S}_Q={{\sigma_c}\over{\kappa_c+\sigma_c}}\Big{\{}\frac{3}{2\sqrt{2}}(\mu^2-1) 
  {J}^2_0 \nonumber
\end{eqnarray}
\vspace{-0.1in}
\begin{eqnarray}
\hspace{0.5in} -
  \sqrt{3}  \mu \sqrt{1-\mu^2} (\cos \chi
  {{\tilde J}}^2_1 + \sin 
  \chi {{\hat J}}^2_1) \nonumber 
\end{eqnarray}
\vspace{-0.2in}
\begin{eqnarray}
\hspace{0.5in} - \frac{\sqrt{3}}{2} (1+\mu^2) (\cos
  2\chi \, {{\tilde J}}^2_2 + \sin 2\chi \, {{\hat J}}^2_2) \Big{\}},
\end{eqnarray}
and

\begin{eqnarray}
S_U={{\sigma_c}\over{\kappa_c+\sigma_c}}\sqrt{3} \,\Big{\{} \sqrt{1-\mu^2} ( \sin \chi
  {{\tilde J}}^2_1 - \cos \chi {{\hat J}}^2_1) \nonumber 
\end{eqnarray}
\vspace{-0.2in}
\begin{eqnarray}
\hspace{1.2in}  +
  \mu (\sin 2\chi \, {{\tilde J}}^2_2 - \cos 2\chi \, {{\hat J}}^2_2) \Big{\}},
\end{eqnarray}
where the orientation of the ray is specified by
$\mu={\rm cos}\theta$ (with $\theta$ the polar angle) and by the azimuthal angle $\chi$. In these source function expressions the ${{J}}^K_Q$ quantities (with $K=0,2$ and $Q=0,1,2$) are the spherical components of the radiation field tensor (see \S~5.11 in Landi Degl'Innocenti \& Landolfi 2004), which are given by the following angular averages of the Stokes parameters (e.g., Manso Sainz \& Trujillo Bueno 1999):

\begin{eqnarray}
  J^0_0=\oint \frac{{\rm d}
  \vec{\Omega}}{4\pi}\,I_{\nu \vec{\Omega}}
\end{eqnarray}
\vspace{-0.3in}
\begin{eqnarray}
  J^2_0=\oint \frac{{\rm d} \vec{\Omega}}{4\pi}
  \frac{1}{2\sqrt{2}} [(3\mu^2-1)I_{\nu \vec{\Omega}}+3(\mu^2-1)Q_{\nu
  \vec{\Omega}}]
\end{eqnarray}
\vspace{-0.3in}
\begin{eqnarray}
  J^2_1=\oint \frac{{\rm d} \vec{\Omega}}{4\pi}
  \frac{\sqrt{3}}{2} e^{i\chi} \sqrt{1-\mu^2}[-\mu(I_{\nu \vec{\Omega}}+Q_{\nu
  \vec{\Omega}})-iU_{\nu \vec{\Omega}}]
\end{eqnarray}
\vspace{-0.3in}
\begin{eqnarray}
  J^2_2=\oint \frac{{\rm d} \vec{\Omega}}{4\pi}
  \frac{\sqrt{3}}{2} e^{2i\chi}
  [\frac{1}{2}(1-\mu^2)I_{\nu \vec{\Omega}}-\frac{1}{2}(1+\mu^2)Q_{\nu
  \vec{\Omega}}-i\mu U_{\nu \vec{\Omega}}] 
\end{eqnarray}
where $I_{\nu\vec{\Omega}}$, $Q_{\nu\vec{\Omega}}$ and $U_{\nu\vec{\Omega}}$ are the Stokes parameters relative to the direction $\vec{\Omega}$ specified by the polar angle $\theta$ and the azimuthal angle $\chi$, with $\mu=\cos \theta$ and $\nu$ the frequency. In these equations, the reference direction for the Stokes $Q$ and $U$ parameters is situated in the plane perpendicular to $\vec{\Omega}$ and lies in the plane containing $\vec{\Omega}$ and the Z-axis of the reference system. 
Note that ${{J}}^2_Q$ (with $Q=1,2$) are complex quantities and that in Eqs. (5)-(7) $\tilde{J}{}^{2}_{Q}$ indicate the {\em real} parts, while $\hat{J}{}^{2}_{Q}$ the {\em imaginary} parts. Note that ${J}{}^{0}_{0}$ quantifies the mean intensity of the continuum radiation, ${J}{}^{2}_{0}$ its  ``anisotropy", and ${J}{}^{2}_{Q}$ (with $Q=1,2$) the breaking of the axial symmetry of the continuum radiation field through the complex azimuthal exponentials that appear inside the angular integrals (i.e., $e^{i\chi}$ for ${J}{}^{2}_{1}$ and $e^{i2\chi}$ for ${J}{}^{2}_{2}$). Obviously, ${J}{}^{2}_{1}$ and ${J}{}^{2}_{2}$ are zero in a plane-parallel or spherically symmetric model atmosphere, but they can have significant positive and negative values in a 3D model (see \S4 below). It is also very important to point out that Eq. (5) is the correct expression for $S_I$ in a 3D medium, and that the approximation that is normally  made for calculating the continuum intensity [i.e., $S_I=B_{\nu}(T)$] is justified only when the effective polarizability is negligible (i.e., when $\sigma_c/(\kappa_c+\sigma_c){\approx}0$).  

Our numerical solution of this physical problem is based on the iterative methods developed by Trujillo Bueno \& Manso Sainz (1999), which require calculating the radiation field tensors at each iterative step. To this end, we have solved Eqs. (4) by applying the 3D formal solver of Fabiani Bendicho \& Trujillo Bueno (1999). The iterative solution of Eqs. (4)-(7) gives us the converged values of the source function components, $S_X$ (with $X=I,Q,U$). Since in the lower solar photosphere the effective polarizability is small (see \S5 below) the well-known $\Lambda$-iteration method of solution can also be safely applied if a sizable number of iterations are performed.  With the converged $S_X$ values, the formal solution of the radiative transfer Eq. (4) allows us to calculate the emergent Stokes parameters $I$, $Q$, $U$ for any desired LOS. 

\section{The atmospheric models and their absolute intensities}

The main solar atmospheric model we have chosen for this investigation is a 3D snapshot model of the solar photosphere, which we have taken from Asplund et al.'s (2000) radiative hydrodynamic simulations of solar surface convection (hereafter, the 3D model). The dashed and dashed-dotted lines of Fig. 1 give the height variation of the kinetic temperature above typical granular and intergranular points in such a 3D model, respectively. In addition, we have used also the semi-empirical 1D model of the ``quiet" Sun atmosphere tabulated in Table II of Maltby et al. (1986) (hereafter, the  MACKKL model), whose thermal structure is given by the solid line of Fig. 1. The model with the label ``${\langle 1D \rangle}$" results when the physical conditions at each height in the 3D model are spatially averaged along the corresponding horizontal plane (see also figure 1 of Stein \& Nordlund 1998).  

There are certainly other 3D models of the solar photosphere we could have chosen, like those resulting from recent magnetohydrodynamic (MHD) simulations (e.g., V\"ogler \& Sch\"ussler 2007). However, the purely hydrodynamical 3D model we have selected suffices for reaching the main objective of this paper, namely to formulate and solve in a ``realistic" model of the solar photosphere the 3D radiative transfer problem of the polarization of the Sun's continuous spectrum. In forthcoming publications we will show the solution of scattering polarization problems in MHD models of the solar photosphere, such as those resulting from the surface dynamo simulations by V\"ogler \& Sch\"ussler (2007).

On the one hand, there are several indications that the 3D solar photospheric model of Asplund et al. (2000) provides a fairly good representation of the thermal and dynamic structure of the quietest regions of the solar photosphere. For example, the synthetic spatially and temporally averaged line profiles of many iron lines show excellent agreement with the observed profiles (e.g., Stein \& Nordlund 2000; Asplund et al. 2000). Moreover, a comparison of spatially resolved observations of the Fe {\sc i} 709.0 nm line with spectral synthesis in a time series of 3D snapshots from Asplund et al.'s (2000) simulations showed also an excellent agreement (Cauzzi et al. 2006). It is important to emphasize that no tunable parameters, such as micro- or macroturbulence, besides the abundance of iron were used in such LTE spectral syntheses, and that a determination of the iron abundance via a detailed comparison of some observables with non-LTE modeling of the solar iron spectrum led us to obtain the meteoritic iron abundance (Shchukina \& Trujillo Bueno 2001). On the other hand, we must also mention that there are some hints that the 3D solar photosphere model of Asplund et al. (2000) is not as accurate as needed for a fully reliable determination of the abundances of some chemical elements, such as that of oxygen.  For example, Ayres et al. (2006) argue that the temperature structure of the 3D model close to the continuum forming layers is slightly too steep, which is in line with an earlier (tentative) conclusion by Asplund et al. (1999) obtained from small discrepancies in the behavior of the wings of the hydrogen Balmer lines and in the behavior of the center-limb variation of the continuum radiation. Moreover, the recent modeling by Shchukina \& Trujillo Bueno (2009) of the scattering polarization observed in the lines of multiplet 42 of Ti {\sc i} suggests that the 3D model of Asplund et al. (2000) is too cool above a height of 400 km, in agreement with the possibility that the temperature gradients of the continuum forming layers could be slightly too strong. 

It is important to emphasize that the input parameters of the solar surface convection simulations of Stein \& Nordlund (1998) and of Asplund et al. (2000) are the surface gravity, the metallicity and the entropy of the inflowing material at the bottom boundary. Therefore, the effective temperature of the simulation is a property which depends on the entropy structure and evolves with time around its mean value following changes in the granulation pattern (cf. Asplund 2007). For this reason, the absolute continuum intensities calculated in a single snapshot of the time-dependent simulation do not have to coincide exactly with those observed. We have calculated the vertically emergent continuum intensities by solving the radiative transfer problem using Eq. (5) for $S_I$ (i.e.,  including the scattering contribution to the continuum source function), and the UV haze opacity strategy described in Bruls, Rutten \& Shchukina (1992) (which applies to wavelengths $\lambda{<}4200$ \AA). For example, the solid line of Fig. 2 shows that for $\lambda{>}4200$ \AA\ the absolute continuum intensities in the 3D snapshot model we have used are slightly smaller than the empirical values. In contrasts, as shown by the dashed line of Fig. 2, for $\lambda{>}4200$ \AA\  the MACKKL model gives absolute continuum intensities virtually identical to those observed in the Sun. We point out that it is natural that the MACKKL model produces such an extraordinary agreement for such absolute continuum intensities, since it is a semi-empirical 1D model based on continuum observations. For $\lambda{<}4200$ \AA\ the measured absolute continuum intensities are only slightly smaller than the intensities calculated in both, the 3D model and in the MACKKL model, which indicates that the haze opacity strategy we have applied is fairly good. It is also logical that the model labelled ${\langle 1D \rangle}$ (which we obtained by horizontally averaging the physical quantities of the 3D model at each height) strongly underestimates the absolute continuum intensities (see the dotted line of Fig. 2). In order to extract from the 3D model a 1D one capable of reproducing the observed continuum intensities we would have to average the physical quantities of the 3D model over surfaces of constant radial monochromatic continuum optical depth $\tau_{\lambda}$ (see, e.g., the mean model that Asplund et al. 2004 show in their fig. 1, which they obtained for $\lambda={5000}$ \AA). This is precisely one of the ``${\langle 1D \rangle}$" models used by Ayres et al. (2006) to argue that the temperature gradients of Asplund's et al. (2000) 3D model is slightly too steep in the continuum formation layers. We have however decided not to use here such a mean model, given that one could actually build many different of such ``${\langle 1D \rangle}$" models by simply changing the wavelength chosen for defining $\tau_{\lambda}$. Another reason is that, in contrast with the radiative transfer approximations on which the work of Ayres et al. (2006) is based, our approach is a fully 3D one --that is, it consists in computing the spatially averaged emergent radiation after calculating the Stokes parameters at each surface point of the chosen 3D snapshot model by solving the radiative transfer equation along each ray with the 3D formal solver of Fabiani Bendicho \& Trujillo Bueno (1999), even along those with $\mu{<}1$ for which our approach takes fully into account the geometrical interaction along the slanted ray of granular and intergranular plasma.

\section{The radiation field tensors of the solar continuous radiation}

As seen in Eqs. (5)-(7) the radiation field tensors defined by equations (8)-(11) play a fundamental role in scattering polarization. As mentioned in \S2, ${J}{}^{0}_{0}$ is nothing but the mean intensity at the spatial point and wavelength under consideration, while ${J}{}^{2}_{0}$ quantifies the ``anisotropy" of the continuum radiation. Since the term of Eq. (9) containing Stokes $I$ makes the dominant contribution, it is clear that ${J}{}^{2}_{0}{>}0$ (${J}{}^{2}_{0}{<}0$) if the intensity of the ``vertical" rays (i.e., those with $|{\mu}|>1/\sqrt{3}$) is larger (smaller) than the intensity of the ``horizontal" rays (i.e., those with 
$|{\mu}|<1/\sqrt{3}$). The quantities ${J}{}^{2}_{Q}$ (with $Q=1,2$) quantify the breaking of the axial symmetry of the continuum radiation field [see Manso Sainz \& Trujillo Bueno (1999) and Manso Sainz (2002) for a detailed, basic radiative transfer investigation of scattering line polarization and the Hanle effect in horizontally inhomogeneous atmospheres]. 

Interestingly, while ${J}{}^{2}_{1}$ and ${J}{}^{2}_{2}$ are zero in a plane-parallel or spherically symmetric model atmosphere, they turn out to fluctuate horizontally at each height in the 3D hydrodynamical model of the solar photosphere, with sizable positive and negative values mainly around the boundaries between the granular and intergranular regions. This can be seen clearly in Fig. 3, which for $\lambda=4600$ \AA\ shows the height variation of the radiation field tensors for all the horizontal grid points of the 3D model. Interestingly, these non-zero values for the real and/or imaginary parts of ${{J}^2_1}$ and ${{J}^2_2}$ imply non-zero values for ${{S}_U}$ (see Eq. 7) and a modification of ${{S}_Q}$ with respect to the approximate case in which only the ${{J}^2_0}$ contribution is retained in Eq. (6). Note also that Eqs. (6) and (7) indicate that the main observable effects of the breaking of the axial symmetry of the continuum radiation field would be non-zero Stokes $Q/I$ and $U/I$ signals at the solar disk center ($\mu=1$) and non-zero Stokes $U/I$ signals at any position off the solar disk center.

It is also of interest to note that since the spatial horizontal averages of the real and imaginary parts of ${{J}^2_1}$ and ${{J}^2_2}$ tend to vanish (see Fig. 3), an approximate expression for estimating the amplitude of the spatially-averaged polarization of the emergent continuum radiation is $Q/I{\approx}S_Q/{S}_I$, with $S_Q$ given by the first term of Eq. (6) and $S_I{\approx}B_{\nu}(T)$. Choosing the reference direction for Stokes $Q$ along the perpendicular to the solar radius vector through the observed point, we obtain

\begin{equation}
{{Q}\over{I}}\,{\approx}\,{{3}\over{2\sqrt{2}}}(1-\mu^2){{\sigma_c}\over{\kappa_c+\sigma_c}}\,{\alpha}\,{{J^2_0}\over{J^0_0}},
\end{equation}
where ${J^0_0}={\alpha}{B_{\nu}}$, with ${\alpha}$ a coefficient of order unity (typically slightly larger than 1) whose horizontally averaged value depends on the height in the 3D atmospheric  model.

\section{The effective polarizability in various atmospheric models}

As seen in Eqs. (6) and (7), the smaller the effective polarizability $\sigma_c/(\kappa_c+\sigma_c)$ the smaller the continuum polarization of the emergent radiation, and the closer to $B_{\nu}(T)$ is $S_I$ (see Eq. 5). Therefore, it is useful to show first the height and wavelength variation of $\sigma_c/(\kappa_c+\sigma_c)$ in some atmospheric models. Fig. 4 shows the height variation of $\sigma_c/(\kappa_c+\sigma_c)$ at 4000 \AA\ in each of the atmospheric models of Fig. 1. Note that below 200 km the largest values are found in the intergranular plasma of the considered 3D hydrodynamical model, and that such values are significant (i.e., of the order of $1\%$ at 4000 \AA, and larger at shorter wavelengths).

As shown above, Eq. (12) provides a suitable formula for {\em estimating} the amplitude of the spatially-averaged polarization of the emergent continuum radiation. Therefore, another worthwhile piece of information is that contained in Fig. 5, which shows the wavelength variation of $\sigma_c/(\kappa_c+\sigma_c)$ at the atmospheric height, $h(\mu=0.1)$, where the monochromatic optical depth [$\tau(\lambda)$] is unity along a LOS with $\mu=0.1$. Note that at such heights there is no significant hint of the Balmer jump in $\sigma_c/(\kappa_c+\sigma_c)$. This will be of interest for our discussion below in \S7.

Figure 6 gives information on the atmospheric heights where $\tau(\lambda)=1$ along a LOS with $\mu=0.1$. Note that in the visible part of the spectrum the larger the wavelength the larger $h(\mu=0.1)$ (because the H$^{-}$ cross-section peaks at 8500 \AA), and that between 3000 \AA\ and 7000 \AA\ the largest height $h(\mu=0.1)$ lies below 200 km in the MACKKL model. We point out that $h(\mu=0.1)$ increases rapidly below 3000 \AA, because in the UV part of the solar spectrum the other continuum opacity contributors play an increasingly important role.

\section{The center-limb variation of the Sun's continuous spectrum}

As pointed out by Milne (1930) the darkening of the Sun's disc towards the limb is a consequence simply and solely of the temperature gradient in the solar atmosphere. Therefore, it is obvious that a careful confrontation between the observed and calculated center-limb variation as a function of wavelength (e.g., between 3000 \AA\ and 7000 \AA) should be considered as one of the key tests of any atmospheric model. In fact, Asplund, Nordlund \& Trampedach (1999) did this exercise for 
time sequences of their 3D solar surface convection simulations by assuming $S_I=B_{\nu}$ (i.e., neglecting the scattering contribution to the continuum source function) and concluded that the theoretical limb-darkening is ``slightly too strong in the UV-blue". This result and their 3D synthesis of the wings of the Balmer lines (which are also sensitive probes of the thermal structure in the lower photosphere) indicated that the temperature structure of their (1999) hydrodynamical model close to the continuum forming layers could be slightly too steep, a possibility that is now being taken seriously by  Ayres et al. (2006) in their paper on the ``solar oxygen crisis".

One important reason for considering here the limb-darkening test is because all previous calculations have assumed $S_I=B_{\nu}(T)$, while ours is based on Eq. (5). Another one is that the stronger the limb-darkening, the larger the anisotropy of the continuum radiation (i.e., the $J^2_0$ tensor of Eq. 9), and the larger the continuum polarization amplitude. For these reasons, we have compared the calculated center-limb variation of the emergent Stokes $I$ parameter with the empirical solar limb-darkening data of Pierce \& Slaughter (1977) and Neckel \& Labs (1994). To this end, for each chosen $\mu$ value and wavelength point between 3000 \AA\ and 7000 \AA, we have computed the spatially averaged intensity after calculating the individual emergent intensities at each surface point of the 3D photospheric model. We emphasize that we have done this by applying the 3D formal solver of Fabiani Bendicho \& Trujillo Bueno (1999) --that is, taking fully into account the geometrical interaction along the slanted ray of granular and intergranular plasma. Figure 7 shows the results of this comparison for 9 selected $\mu$-values. As it can be seen, the agreement between the calculated and observed limb darkening laws is impressive, though the center-limb variation is slightly too strong at wavelengths around $\lambda=5000$ \AA\ and for $\mu$-values around $\mu=0.3$. Note also that the more recent observations of Neckel \& Labs (1994) are in very good agreement with those of Pierce \& Slaughter (1977). For the 1D semi-empirical model of Maltby et al. (1986) we find that the agreement between the calculated and the observed limb-darkening functions is fairly good towards the red part of the spectrum, but the errors are rather significant at blue wavelengths.

It is very important to emphasize again that the calculated intensities of Fig. 7 were obtained by solving the radiative transfer Eq. (4) using Eq. (5) for $S_I$ (i.e., including the scattering contribution to the continuum source function). However, many investigations assume instead $S_I=B_{\nu}(T)$ (i.e., $\sigma_c=0$ in Eq. 5), which according to Fig. 4 is a questionable approximation, especially in the intergranular plasma and for the shortest wavelengths of the spectral range considered in this paper. Fig. 8 shows, for $\lambda=3000$ \AA, the limb darkening functions calculated in the 3D model assuming $\sigma_c=0$ and $\sigma_c{\ne}0$. In conclusion, the assumption $S_I=B_{\nu}(T)$ (i.e., $\sigma_c=0$ in Eq. 5) leads to a small but significant error in the calculated center-limb variation, and should therefore be avoided for a reliable evaluation of the realism of atmospheric models at continuum wavelengths sensibly smaller than 5000 \AA. The consideration of the scattering contribution in the $S_I$ expression may be also important for a more accurate calculation of the radiative flux divergence at each time step of magneto-hydrodynamic simulations. In fact, it is expected to yield smoother temperature gradients in the cooling surface layer (see Skartlien 2000). This might be important for a more accurate inference of stellar chemical abundances through the use of 3D atmospheric models resulting from stellar surface convection simulations.   

\section{The RMS contrast of solar granulation}

The root-mean-square (RMS) contrast of 
solar granulation is the standard deviation of the continuum intensity at disk center, normalized to the spatially averaged intensity. Probably, this is the easiest statistical parameter that can be used to estimate the reliability of the horizontal fluctuations of a 3D model. 

The solid curve of Fig. 9 shows the results of our calculations of the wavelength variation of the RMS contrast of the emergent continuum intensity in the 3D photospheric model. To this end, we solved the radiative transfer problem explained in \S2 using Eq. (5) for $S_I$. The dotted curve shows what happens when the synthetic monochromatic images are degraded so as to mimic the effects of the solar space telescope Hinode. To this end, we have taken into account the same effects considered by Danilovic et al. (2008), but without including any defocus in Fig. 9. We have done this by applying an image degradation code kindly provided to us by Dr. J. A. Bonet (IAC). Interestingly, at the 6301 \AA\  continuum wavelength observed by the Hinode spectropolarimeter (see the quiet Sun observations of Lites et al. 2008) the RMS contrast found in a small area of $201{\times}201$ pixels contained in the observed field of view is 7.8\%, while the RMS contrast calculated using  the vertically emergent intensities from the 3D model after applying the above-mentioned image degradation is 8.6\%. Such a small area of the full field of view observed by Lites et al. (2008) was selected because it was one of the regions showing the weakest circular polarization signals. In principle, the ensuing small discrepancy between the calculated and the observed contrasts at the chosen 6301 \AA\ continuum wavelength can be ascribed either to the same effects pointed out by Danilovic et al. (2008) (e.g., a slight defocus and/or straylight and slight imperfections of the instrument), and/or to a plausible impact of the spatially unresolved magnetic field that Trujillo Bueno et al. (2004) inferred via a joint analysis of the Hanle effect in atomic and molecular lines, and/or to the fact that in all current simulations of solar surface convection the radiative flux divergence is calculated assuming that the scattering coefficient is zero.

The degree of realism of the horizontal fluctuations in the 3D model can be investigated better by contrasting the observed and the calculated histograms of the ${\lambda}6301$ continuum intensity distribution (see also Stein \& Nordlund 2000). Fig. 10 shows a comparison of the histogram calculated in the 3D snapshot model with that corresponding to the observed intensity fluctuations in the above-mentioned small area of $201{\times}201$ pixels contained in the field of view of Lites et al's (2008) quiet Sun observations. Note that the double-peaked distribution corresponding to the raw simulation (see the thin solid line) practically disappears when the smearing effect of the {\em Hinode} telescope is taken into account (see the dotted line), and that the RMS contrast decreases from 8.6\% (left panel) to 7.65\% (right panel) when the defocusing of the telescope varies from zero to 1.5 mm. At first sight,  
the small discrepancy seen in Fig. 10 between the calculated (dotted lines) and the observed (thick solid lines) histograms suggests that there could be a slight excess of (dark) intergranular regions in the 3D snapshot model we obtained from Asplund et al's. (2000) simulations. We point out, however, that 
such a small discrepancy could be simply due to straylight effects caused by slight imperfections of the observing instrument.

\section{The anisotropy factor}

The anisotropy factor is a fundamental quantity for scattering polarization. Its definition is (e.g., Landi Degl'Innocenti \& Landolfi 2004)

\begin{equation}
w=\sqrt{2}{{J^2_0}\over{J^0_0}},
\end{equation}
where $J^0_0$ is the mean intensity (see Eq. 8) and $J^2_0$ the
radiation field tensor of Eq. (9) which quantifies basically the degree of ``vertical" vs. ``horizontal" anisotropy of the radiation field under consideration (i.e., the solar continuum radiation). As shown below, the anisotropy factor is another quantity that can be used for obtaining some insight on the reliability of any given atmospheric model.

Note that the tensor $J^2_0$ is dominated by the Stokes $I$ contribution of Eq. (9), because $Q<<I$. Therefore, we can easily use the center-limb variation of the {\em observed} solar continuum intensity to obtain the empirical wavelength variation of $w$. To this end, we have used the original data tabulated by Pierce \& Slaughter (1977) in their Table IV and by Neckel \& Labs (1994) in their Table I, instead of the limb darkening data given by Pierce (2000). In this way, we have been able to obtain directly (i.e., without interpolations) the empirical anisotropy factor at more wavelength points, something useful for improving the wavelength resolution around the location of the Balmer jump.

The above-mentioned empirical anisotropy factor has to be compared with that obtained by horizontally averaging the theoretical $w(x,y)$-values corresponding to a given ``optically thin" atmospheric height (e.g., $z=600$ km), since if the atmospheric model is reliable (at least in what concerns its ability to yield accurate values of spatially averaged observables) there should be agreement between the theoretical and empirical anisotropy factors as a function of wavelength. 

Figure 11 shows the empirical and theoretical results for various atmospheric models. The filled circles show the wavelength variation of the empirical anisotropy factor, obtained from the previously mentioned limb darkening observations. The thick solid line (with red color in the electronic version of this paper) shows the above-mentioned theoretical $w$-values obtained by solving the continuum polarization problem in the 3D hydrodynamical model. Interestingly enough, the amplitude of the  Balmer jump in the theoretical $w(\lambda)$ functions is very sensitive to the thermal and density structure of the atmospheric model. This is easy to understand if one takes into account that the anisotropy factor is a sensitive function of the source function gradient (see fig. 4 in the paper by Trujillo Bueno 2001). Note that the best overall agreement is found for the 3D model, when solving consistently the 3D radiative transfer continuum polarization problem as explained in \S2. In particular, the agreement is fairly good for  $\lambda{<}6000$ \AA, including the wavelength region around the Balmer jump, with the obvious exception of the above-mentioned 80 \AA\ shift towards larger wavelengths. However, some small discrepancies are present, especially for wavelengths between 6000 \AA\ and 7000 \AA. In this spectral region the anisotropy factor values calculated in the 1D model of Maltby et al. (1986) seems to be closer to the empirical ones (see the thin solid line of Fig. 11).

\section{The polarization of the Sun's continuous spectrum}

As mentioned in \S1 Stenflo (2005) used Gandorfer's (2000; 2002; 2005) atlas of the second solar spectrum between 3161 \AA\ and 6995 \AA\ to determine the polarization of the Sun's continuous radiation at a disk position with $\mu=0.1$. His determination was {\em semi-empirical} because it was based on a model for the behavior of the spectral lines that depolarize the continuum polarization. His strategy consisted in choosing spectral windows dominated by purely depolarizing lines and in modeling the observed relative line depth in $Q/I$ in terms of the relative line depth in Stokes $I$. 
The key parameter of his model is $\alpha$, whose value determines whether the relative $Q/I$ line depth is similar ($\alpha=1$) or larger ($\alpha{<}1$) than the Stokes $I$ line depth. The two other free parameters of Stenflo's (2005) fitting procedure were $p_c$ (i.e., the continuum polarization) and $p_0$ (i.e., the zero point of the polarization scale). Unfortunately, the quality of the fit could not be used as a good criterion for the selection of $\alpha$, whose value has a small but significant impact on the final $p_c$ value. Nevertheless, after judicious trials with various 3-parameter fits he concluded that the most likely representation of the Sun's continuum polarization as a function of wavelength is that corresponding to $\alpha=0.6$, with an error bar given by the $p_c(\lambda)$ curves corresponding to $\alpha=1$ and $\alpha=0.3$ (see his Fig. 5). The dotted lines of Fig. 12 reproduce his results, with the central curve showing the most likely representation and the two outer ones indicating the approximate lower and upper limits. Note that in Stenflo's (2005) semi-empirical determination the polarized Balmer jump is shifted from the series limit to significantly larger wavelengths, a feature which according to Stenflo (2005) is due to pressure broadening of the Balmer lines from the statistical Stark effect. Note that the central dotted curve of Fig. 12 indicates that the amplitude of the Balmer jump in the semi-empirical $Q/I$ is very significant. It is however important to emphasize that each of the dotted curves of Fig. 12 correspond to a different value of Stenflo's (2005) fitting parameter $\alpha$.

We now turn to contrasting the results of our radiative transfer modeling with the empirically determined continuum polarization. Concerning the 1D quiet Sun model of Maltby et al. (1986) we see in the right panel of Fig. 12 that between 4000 \AA\ and 7000 \AA\ the calculated $Q/I$ lies systematically higher than the semi-empirical values, in agreement with Stenflo's (2005) conclusion that the 1D radiative transfer modeling of Fluri \& Stenflo (1999) for $\lambda{>}4000$ \AA\ seems to overestimate the continuum polarization amplitudes. Note also that the amplitude of the polarized Balmer jump predicted by the MACKKL 1D model is substantially smaller than that seen in the empirical $Q/I$ values. Interestingly enough, as shown by the left panel of Fig. 12, the solution of the continuum polarization problem in the 3D photospheric model is in significantly better agreement with Stenflo's (2005) semi-empirical determination. However, some small discrepancies remain. Obviously, our modeling shows that the polarized Balmer jump is located at the exact wavelength of the Balmer limit, since in this first investigation we have not included the physical ingredient that is thought to be responsible of its shift towards larger wavelengths. Of particular interest is the fact that the amplitude of the Balmer jump in the $Q/I$ calculated in the 3D model is twice larger than that found when using the 1D model of Maltby et al. (1986), but it is also noteworthy that this enhanced amplitude is still slightly smaller than that seen in Stenflo's (2005) semi-empirical determination, as given by the central dotted curve of Fig. 12. Of additional interest is that the slope of the theoretical $Q/I$ curves at wavelengths smaller than that of the Balmer limit is significantly smaller than that of the semi-empirical curves. It would be worthwhile to investigate the main physical reason for these small but significant discrepancies. One possibility is that they are caused by small errors in the thermal and/or density structure of the 3D model, as a result of missing physical ingredients (e.g., the impact of a ``hidden" (unresolved) magnetic field and/or the currently used LTE approximation for calculating the radiative flux divergence at each time step of the hydrodynamic simulation). Another, more likely possibility, is that they are due to errors in Stenflo's (2005) semi-empirical determination of the polarization of the solar continuous spectrum, because it was based on a model for the behavior of the depolarizing lines which required the use of a fitting parameter ($\alpha$) assumed to be independent of wavelength. This last possibility is reinforced by the very good agreement that is however found between the calculated and the empirical values of the anisotropy factor (see Fig. 11).  

Finally, information on the center-limb variation of the polarization of the continuous radiation calculated in the 3D model for  several wavelengths can be seen in Fig. 13.

\section{The effects of symmetry breaking on the continuum polarization}

Figure 14 shows the center-limb variation of the linear polarization of the  continuum radiation at 4600 \AA, which we have calculated by solving the continuum polarization problem in the 3D model. The aim of this figure is to inform the reader about the $Q/I$ and $U/I$ signals that we would see if we could measure with very high spatio-temporal resolution and polarimetric sensitivity the linear polarization in a continuum window close to the wavelength of the Sr {\sc i} line at 4607 \AA. A similar figure, but for the linear polarization produced by scattering processes in the Sr {\sc i} 4607 \AA\ line itself, can be found in Trujillo Bueno \& Shchukina (2007)  (see their Fig. 1), which can be compared directly with Fig. 14 for the (substantially smaller) continuum polarization amplitudes.

As expected, at $\mu=0.1$ and $\mu=0.5$ (see the left and middle top panels of Fig. 14) the $Q/I$ continuum signals are almost everywhere positive, because far away from the solar disk center the term of Eq. (6) proportional to ${{S}^2_0}$ makes the dominant contribution. Note that this term is proportional to ${{J}^2_0}$, which quantifies the anisotropy factor of the radiation field. In any case, it is important to point out that the other terms of Eq. (6) (i.e., those caused by the symmetry breaking effects quantified by the ${{J}^2_1}$ and ${{J}^2_2}$ tensors) do influence the local values of $Q/I$. Note that for $\lambda=4600$ \AA\ $Q/I$ at $\mu=0.1$ varies between about $0.06\%$ and $0.12\%$, while the range of variation at $\mu=0.5$ lies between $0\%$ and $0.025\%$. The spatially averaged 
$Q/I$ amplitude is about $0.09\%$ at $\mu=0.1$ and $0.01\%$ at $\mu=0.5$. The standard deviations ($\sigma$) of the $Q/I$ fluctuations are approximately $0.012\%$ at $\mu=0.1$ and $0.0044\%$ at $\mu=0.5$.

As shown in the corresponding bottom panels of Fig. 14    
the Stokes $U/I$ signals are non-zero, with a typical spatial scale of the fluctuation similar to that of $Q/I$, but with positive and negative values lying between about $-0.01\%$ and $0.01\%$ at $\mu=0.5$ (with a $\sigma{\approx}0.0038\%$)
and between approximately $-0.02\%$ and $0.02\%$ at $\mu=0.1$ (with a $\sigma{\approx}0.0075\%$). Such $U/I$ signals are exclusively due to the symmetry breaking effects caused by the horizontal atmospheric inhomogeneities, which are quantified by the tensors ${J}^2_1$ and ${J}^2_2$. The spatially averaged $U/I$ amplitudes are very small. 

The right panels of Fig. 14 show that the $Q/I$ and $U/I$ signals at the solar disk center ($\mu=1$) have subgranular patterns. In this forward scattering geometry both $Q/I$ and $U/I$ have positive and negative values, which are exclusively due to the symmetry breaking effects (see Eqs. 6 and 7 for $\mu=1$). At 4600 \AA\ such values vary between $-0.006\%$ and $0.006\%$, approximately, with a $\sigma{\approx}0.001\%$). In summary, for visible wavelengths like 4600 \AA\ the spatially averaged $Q/I$ and $U/I$ values at the solar disk center are very small. However, at sufficiently shorter wavelengths the predicted linear polarization amplitudes are much higher and measurable, at least at $\mu=0.5$. For example, for $\lambda=3000$ \AA\ the linear polarization amplitudes are one order of magnitude larger than those seen in Fig. 14.

\section{Concluding comments}

We have formulated and solved the 3D radiative transfer problem of the
polarization of the stellar continuous radiation, which in solar-like atmospheres is principally produced by the interaction between radiation and hydrogen plus free electrons (Rayleigh and Thomson scattering). Of particular interest is that we have taken into account not only the anisotropy of the stellar continuum radiation, but also the symmetry-breaking effects caused by the horizontal atmospheric inhomogeneities produced by the stellar surface convection. As shown in the last figure of this paper, detection of the observational signatures of such symmetry-breaking effects would require observing the linear polarization of the Sun's continuum radiation at high spatial and temporal resolution, so as to be able to see clearly the details of the solar granulation pattern. In particular, at the very center of the solar disk, where we observe the forward scattering case, the predicted $Q/I$ and $U/I$ signals have a subgranular pattern which is solely due to the symmetry breaking effects we have discussed in \S10 (see also Trujillo Bueno \& Shchukina 2007, for the impact of such effects on the scattering polarization of the Sr {\sc i} 4607 \AA\ line).  

Without spatial and/or temporal resolution $U/I{\approx}0$ and the only observable quantity for the solar case is of course $Q/I$, whose wavelength variation at a disk position close to the limb (i.e., at $\mu{\approx}0.1$) was extracted semi-empirically by Stenflo (2005) from Gandorfer's (2000; 2002, 2005) atlas of the second solar spectrum. The fact that collisional and/or magnetic depolarization do not play any role on the linear polarization of the continuum radiation of the Sun's visible spectrum implies that we can use the continuum polarization to evaluate the degree of realism of any given solar photospheric model, such as the 3D snaphot model we have taken from Asplund's et al. (2000) hydrodynamical simulations. Overall, our 3D radiative transfer modeling of the polarization of the solar continuum radiation in such a 3D model shows a notable agreement with Stenflo's (2005) 
semi-empirical data, significantly better than that obtained via the use of 1D atmospheric models. Of especial interest is to note that the amplitude of the Balmer jump in $Q/I$ is larger in 3D than in 1D, which is in line with the sizable amplitude of the Balmer jump seen in Stenflo's (2005) semi-empirical values. 

Given that our 3D modeling of the polarization of the Sun's continuous radiation gives us 
Stokes $I$, in addition to Stokes $Q$ and $U$, we have taken the opportunity of investigating also the reliability of the above-mentioned 3D snapshot model by considering other (more familiar) quantities, like the absolute continuum intensities (Fig. 2), the center-limb variation of the continuum intensity (Fig. 7), the RMS contrast of solar granulation at the 6301 \AA\ continuum wavelength observed with the spectropolarimeter of the {\em Hinode} space telescope (Fig. 9), and histograms of the ensuing continuum intensity fluctuations (Fig. 10). In particular, Fig. 7 contrasts the observed and the calculated wavelength variation of the center-limb variation of the solar continuum intensity. Although the overall agreement is indeed impressive, there are however some small but significant discrepancies in the center-limb curves with $0.2 {\le} {\mu} {\le} 0.5$ and for wavelengths around $\lambda=5000$ \AA, which support the possibility that the temperature gradients in the continuum forming layers of Asplund et al's (2000) 3D model could be {\em slightly} too steep. We emphasize that we have done this type of investigations by using Eq. (5) for $S_I$ --that is, taking into account that the continuum source function is not exactly equal to the Planck function, because the scattering contribution is not completely negligible, especially for wavelengths sensibly smaller than 5000 \AA. In this respect, it is of interest to note that around the visible ``surface" layers the effective polarizability in the downward-moving intergranular lane plasma is significantly larger than in the upward-moving granule plasma, with values of the order of $1\%$ at 4000 \AA\ (see Fig. 4).

The reported small discrepancies between the measured and the calculated observables might be due to the fact that the radiative flux divergence at each time step in Asplund et al's (2000) simulations of solar surface convection was computed assuming $S_I=B_{\nu}(T)$ (i.e., neglecting the scattering contribution to the continuum source function). Another possibility could be a plausible impact of the unresolved magnetic field that Trujillo Bueno et al. (2004; 2006) inferred via a joint analysis of the Hanle effect in the Sr {\sc i} 4607 \AA\ line and in the C$_2$ lines of the Swan system. These authors concluded that the bulk of the ``quiet" solar photosphere is teeming with tangled magnetic fields at subresolution scales, with a mean field strength of the order of 100 gauss, which might be important for the overall energy balance of the solar atmosphere. As shown in this paper, the agreements between the observed and the calculated center-limb variation (see Fig. 7) and between the inferred and the calculated continuum polarization (see the left panel of Fig. 12) are so good that we think that our conclusion on the presence of a substantial amount of ``hidden" magnetic energy in the quiet solar photosphere will remain valid after similar Hanle-effect investigations are carried out using instead the forthcoming, new generation of 3D models (which will probably go beyond the current approach for calculating the radiative flux divergence). Such future models of stellar surface convection and magnetoconvection will be useful for clarifying whether or not the Sun has a sub-solar metallicity, but for further progress in our understanding of the small-scale magnetic activity of our nearest star it is important to perform increasingly realistic solar surface dynamo simulations and to use them as models of the solar photosphere.

\newpage 

\acknowledgments
{\bf Acknowledgments}
We thank Jos\'e Antonio Bonet Navarro and Santiago Vargas (both from the Instituto de Astrof\'\i sica de Canarias, IAC) for allowing us to use their image degradation code for modeling the smearing effect produced by the {\em Hinode} space telescope. Thanks are also due to Andr\'es Asensio Ramos (IAC) for useful discussions about the absolute intensities of the solar continuum radiation and to
H\'ector Socas-Navarro (IAC) for providing the histograms of the solar granulation images observed by {\em Hinode}, which is a Japanese mission developed and launched by ISAS/JAXA with NAOJ as domestic partner and NASA and STFC (UK) as international partners. Finantial support by the Spanish Ministry of Science through project AYA2007-63881 and by the National Academy of Sciences of Ukraine through project 1.4.6/7-233B is also gratefully acknowledged.

%%%%%%%%%%%%%%%%%%%%%%%%%%%%%%%%%%%%%%%%
%%%%%%%%%%%%%%%%%%%%%%%%%%%%%%%%%%%%%%%%
% FIGURES
%%%%%%%%%%%%%%%%%%%%%%%%%%%%%%%%%%%%%%%%
%%%%%%%%%%%%%%%%%%%%%%%%%%%%%%%%%%%%%%%%
\begin{figure}
\plotone{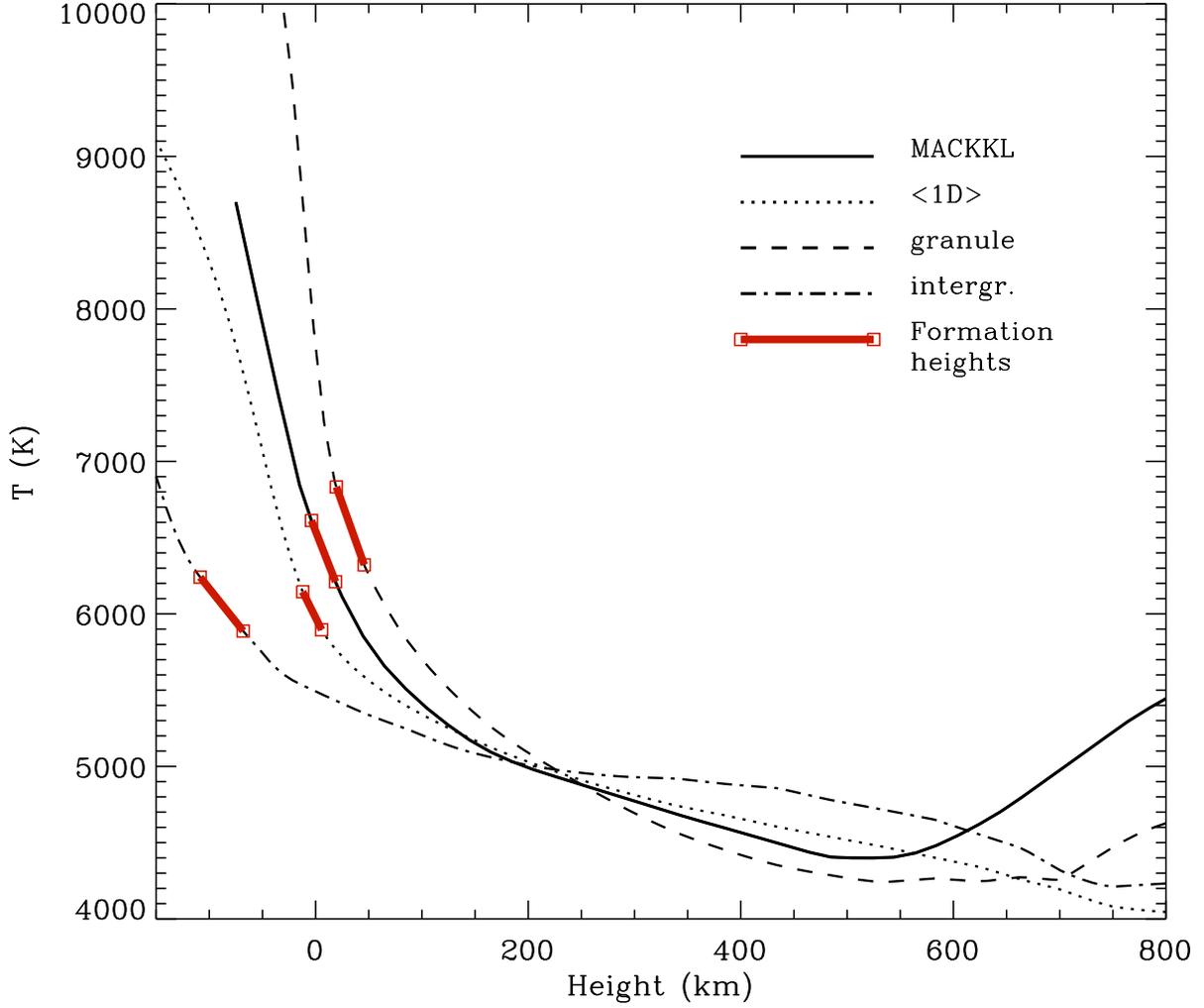}
\caption{Temperature stratification of some atmospheric models. 
Solid line: MACKKL model. The dashed and the dashed-dotted lines are examples of the temperature variation at the center of a granular region and at the center of an intergranular region, respectively, 
which we have extracted from the 3D snapshot model (see the two points indicated in figure 2 of Shchukina \& Trujillo Bueno 2001). Dotted line: the temperature stratification that results when horizontally averaging at each height in the 3D model.    
Note that for each model the figure gives also the range of atmospheric heights where for the continuum radiation 
with 3000 \AA\ ${\le}\,{\lambda}\,{\le}$ 7000 \AA\ the monochromatic optical depth $\tau(\lambda)=1$ for simulated disk-center observations.
\label{fig:atmos}}
\end{figure}

\clearpage 

\begin{figure}
\plotone{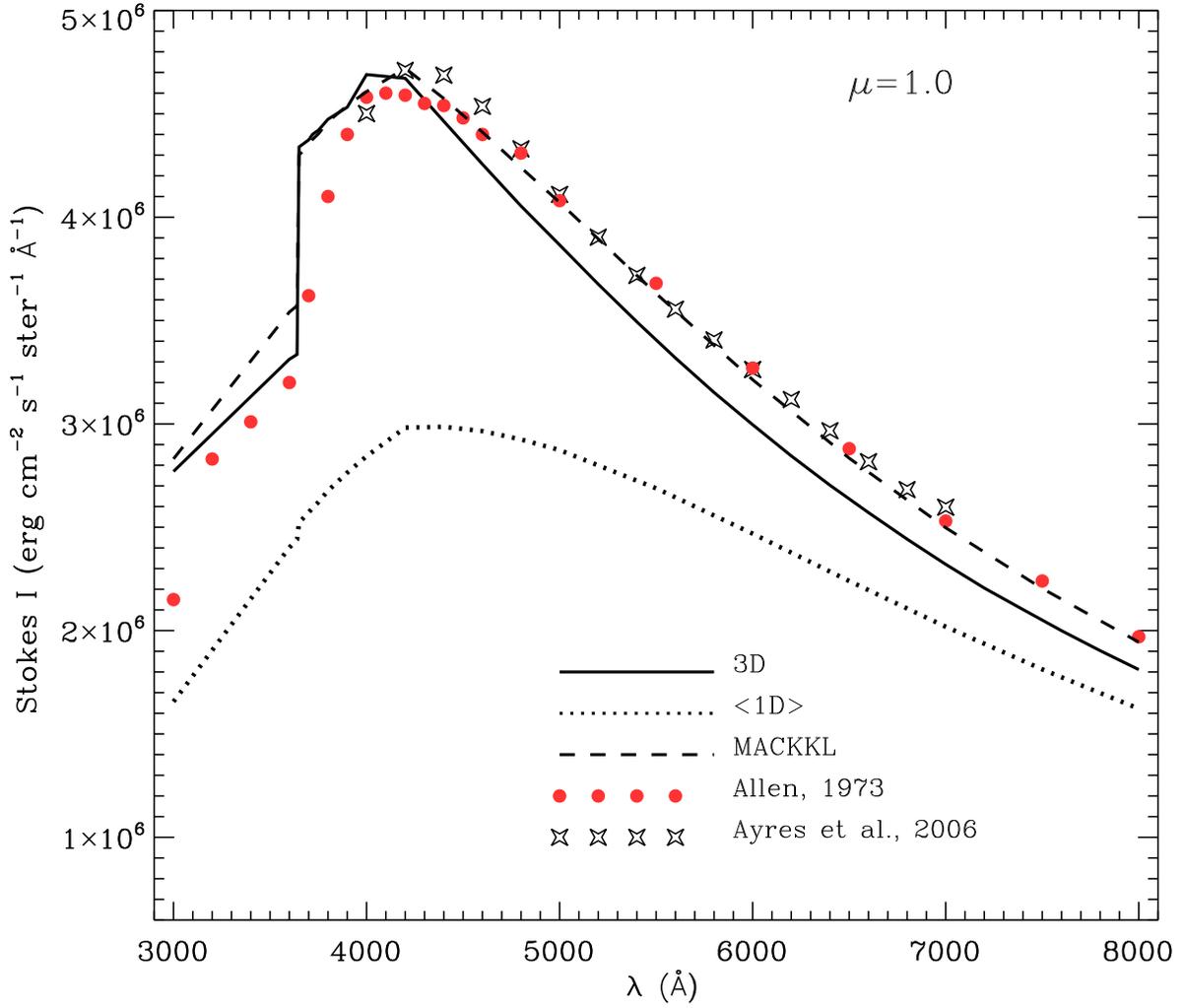}
\caption{The calculated vs. the observed absolute continuum intensities at ther solar disk center. 
\label{fig:absolute}}
\end{figure}

\clearpage 

\begin{figure}
\plotone{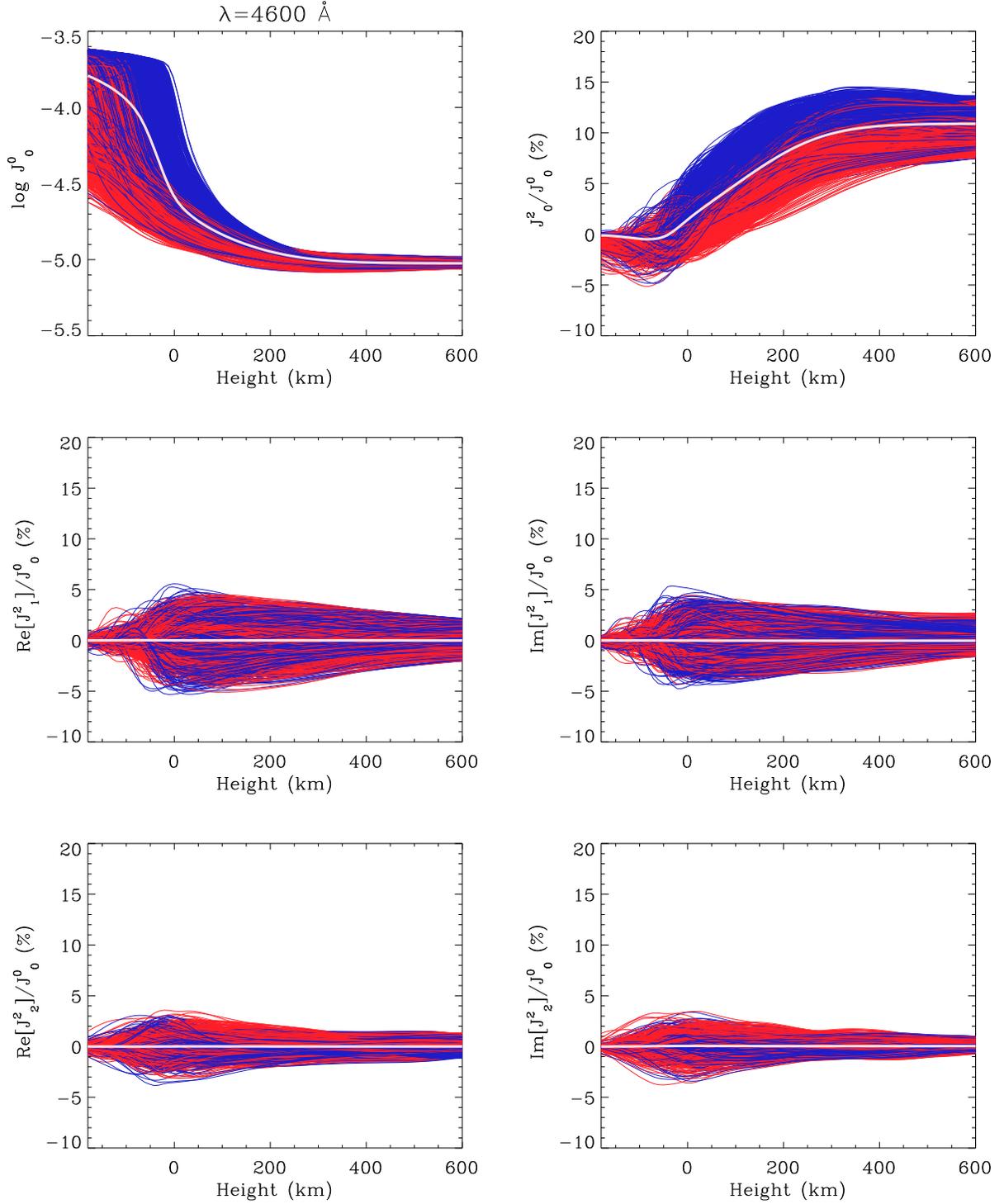}
\caption{For $\lambda=4600$ \AA\ the various panels of this figure show the height variation of $J^0_0$ (see Eq. 8), of 
$J^2_0/J^0_0$  (see Eq. 9) and of the real and imaginary parts of $J^2_Q/J^0_0$ (with $Q=1,2$; see Eqs. 10 and 11) for all the horizontal grid points of the 3D model. In the electronic version of this figure the blue curves correspond to the upflowing regions 
of the 3D model and the red ones to the downflowing ones. 
\label{fig:tensors}}
\end{figure}

\clearpage 

\begin{figure}
\plotone{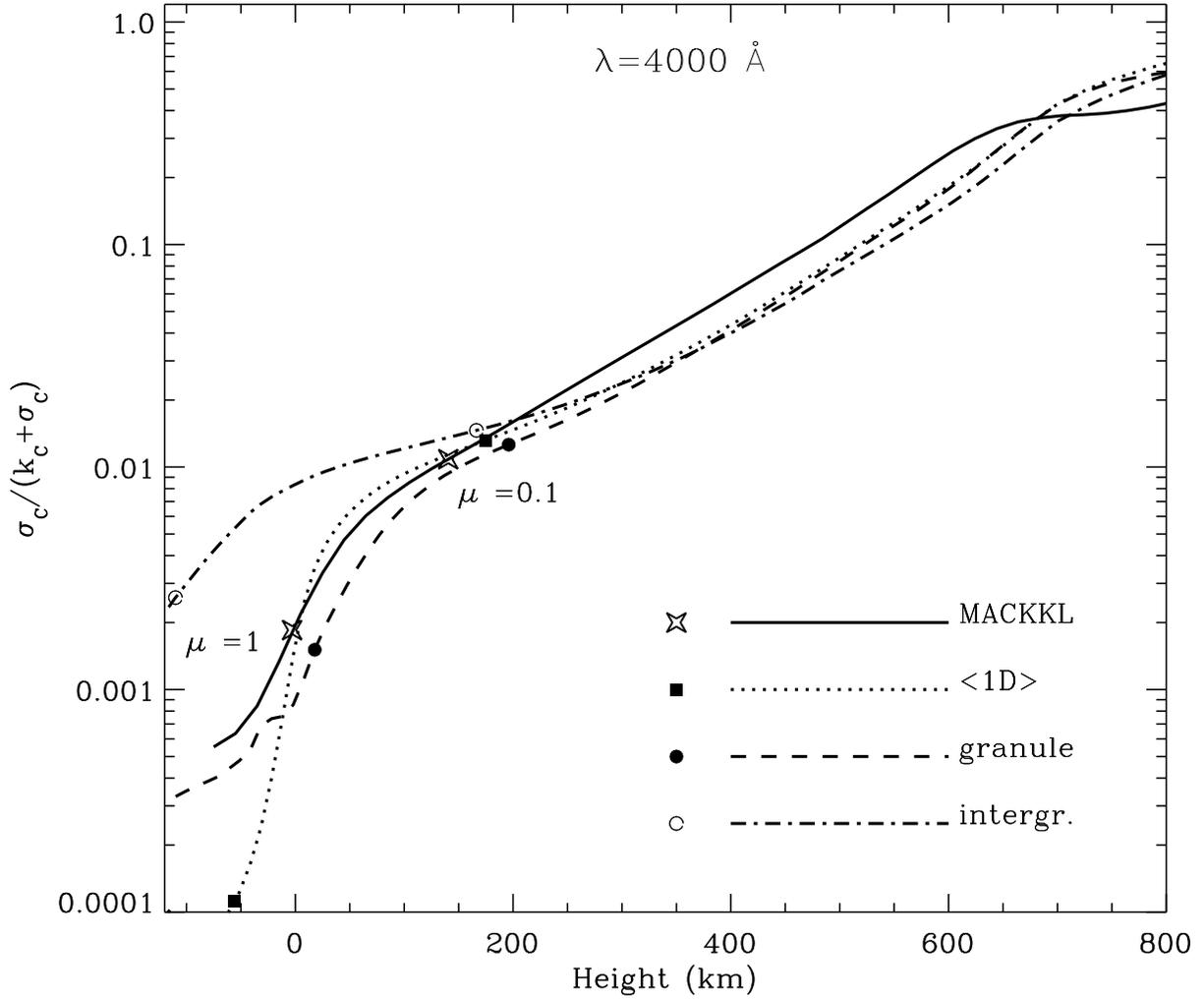}
\caption{For $\lambda=4000$ \AA\ the figure shows the height variation of the effective polarizability, $\sigma_c/(\kappa_c+\sigma_c)$, for each of the atmospheric models of Fig. 1. Note that the two symbols on each curve indicate the atmospheric height where the monochromatic optical depth is unity along the indicated line of sight ($\mu=1$ or $\mu=0.1$).  
\label{fig:polarizability-height}}
\end{figure}

\clearpage 

\begin{figure}
\plotone{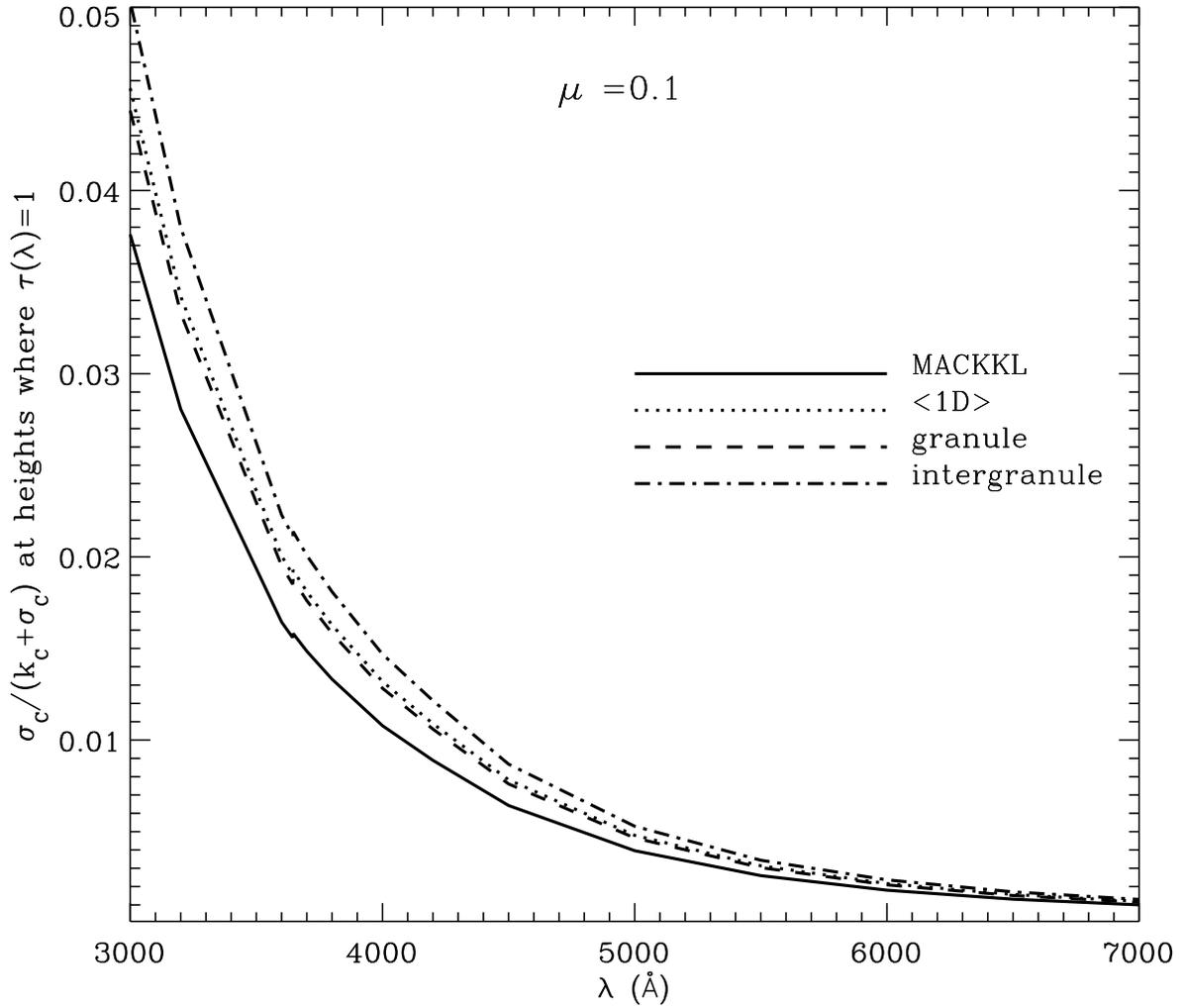}
\caption{The wavelength variation of $\sigma_c/(\kappa_c+\sigma_c)$ at the atmospheric height where the corresponding optical depth $\tau(\lambda)$ is unity along a LOS with $\mu=0.1$. Note that there is no significant hint of the Balmer jump.
\label{fig:polarizability-lambda}}
\end{figure}

\clearpage 

\begin{figure}
\plotone{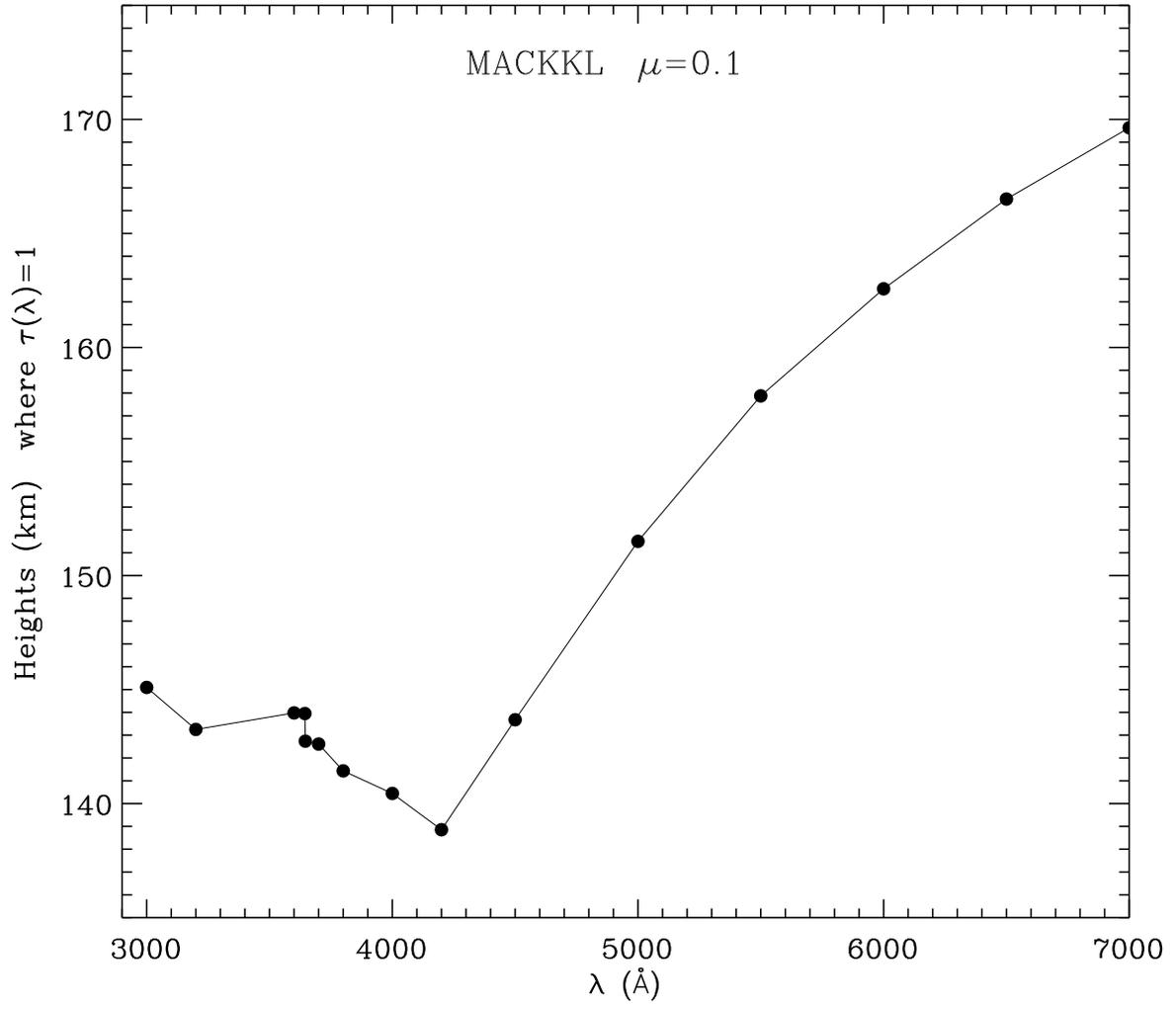}
\caption{The variation with wavelength of the atmospheric height where $\tau(\lambda)=1$ along a LOS with $\mu=0.1$.
\label{fig:height}}
\end{figure}

\clearpage

\begin{figure}
\plotone{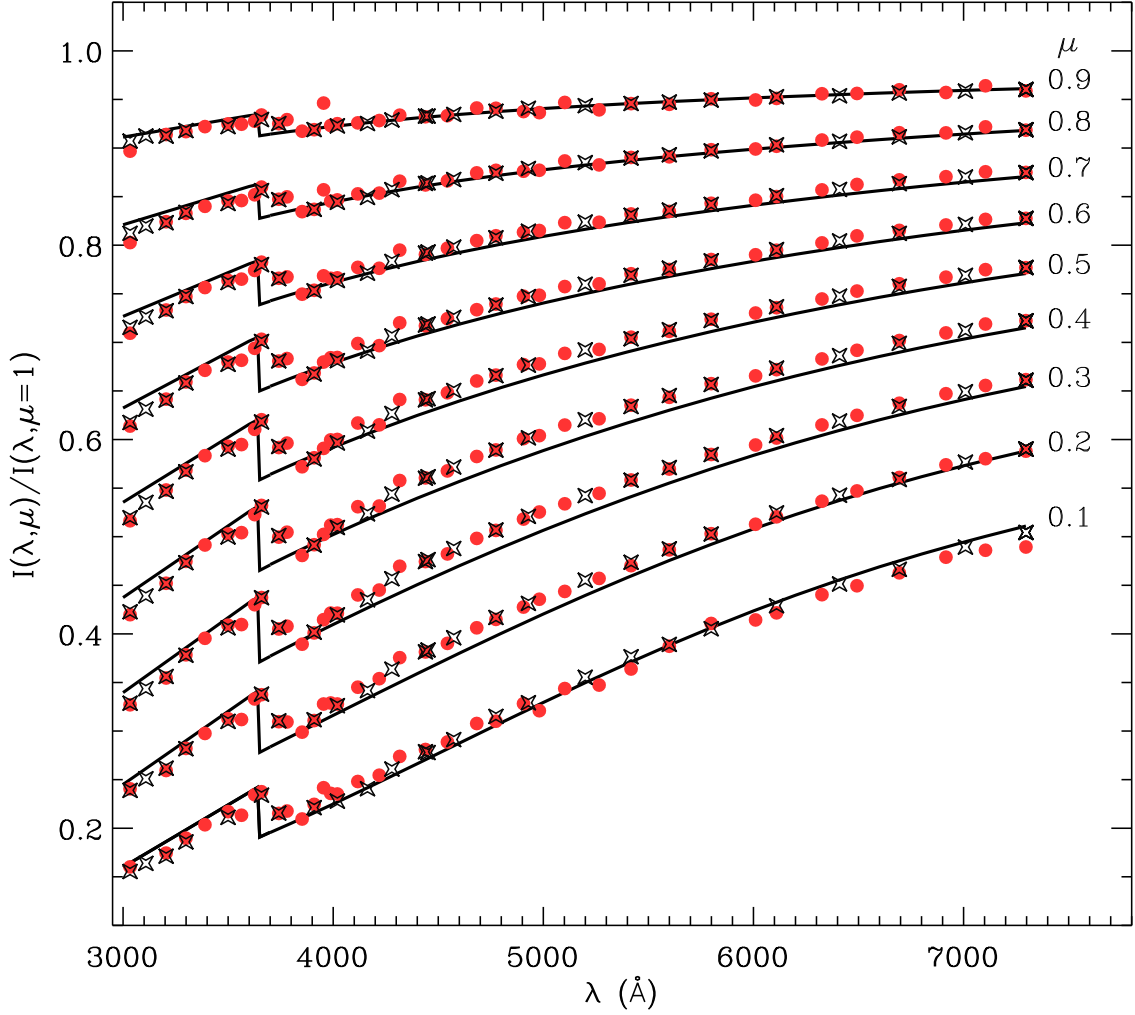}
\caption{The symbols show the variation with wavelength of the observed solar continuum intensity for some selected $\mu$-values, normalized to the  ensuing disk center value. The filled circles correspond to the observations of Pierce \& Slaughter (1977), while the stars to those of Neckel \& Labs (1994). The solid lines show the results of our radiative transfer calculations in the 3D model.
\label{fig:CLV-lambda}}
\end{figure}

\clearpage  

\begin{figure}
\plotone{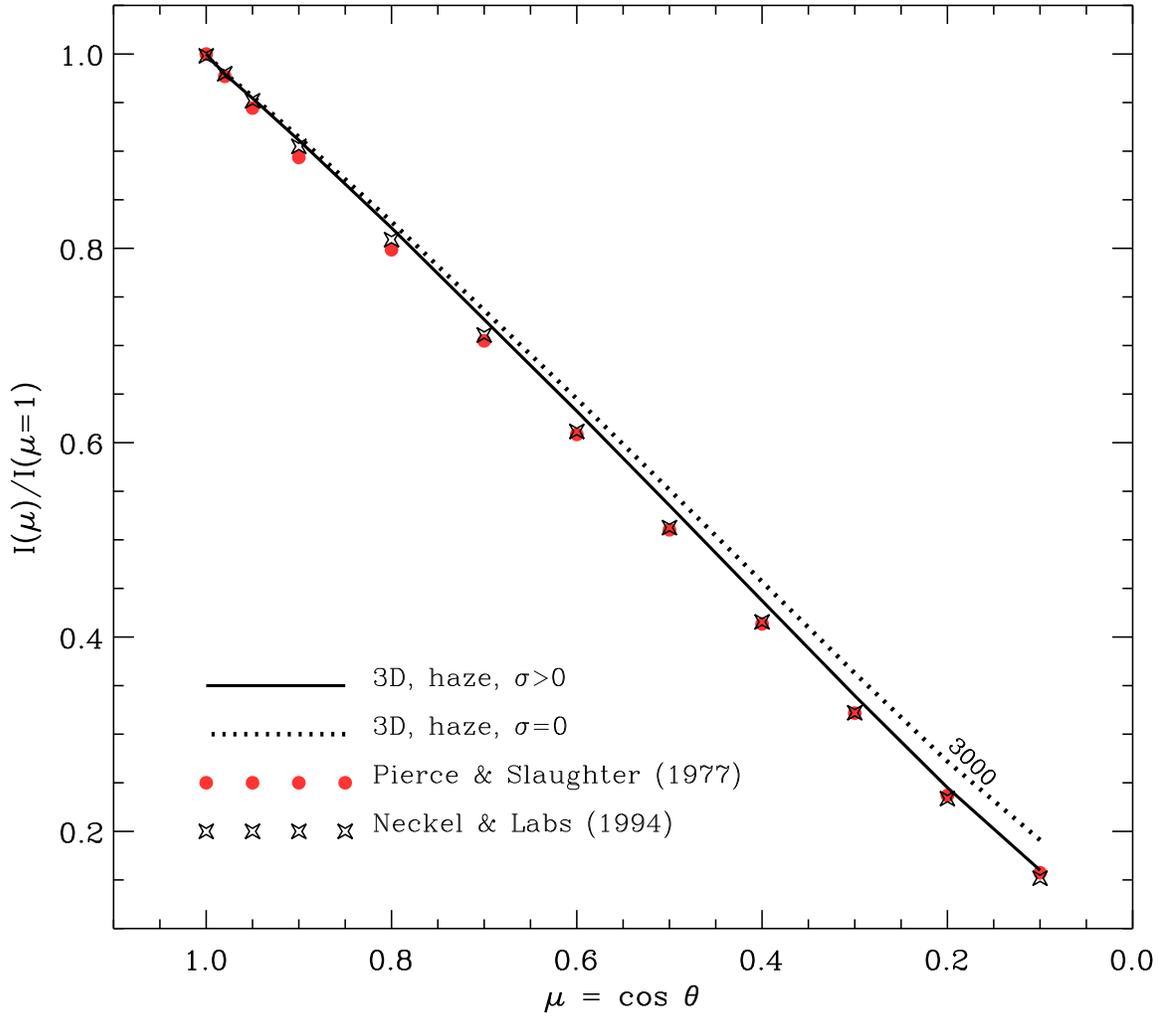}
\caption{The center-limb variation calculated in the 3D model for $\lambda=3000$ \AA\ is compared with the observations, taking into account and neglecting the scattering contribution to the continuum source function. Both calculations accounted for the UV haze opacity.
\label{fig:sigma}}
\end{figure}

\clearpage

\begin{figure}
\plotone{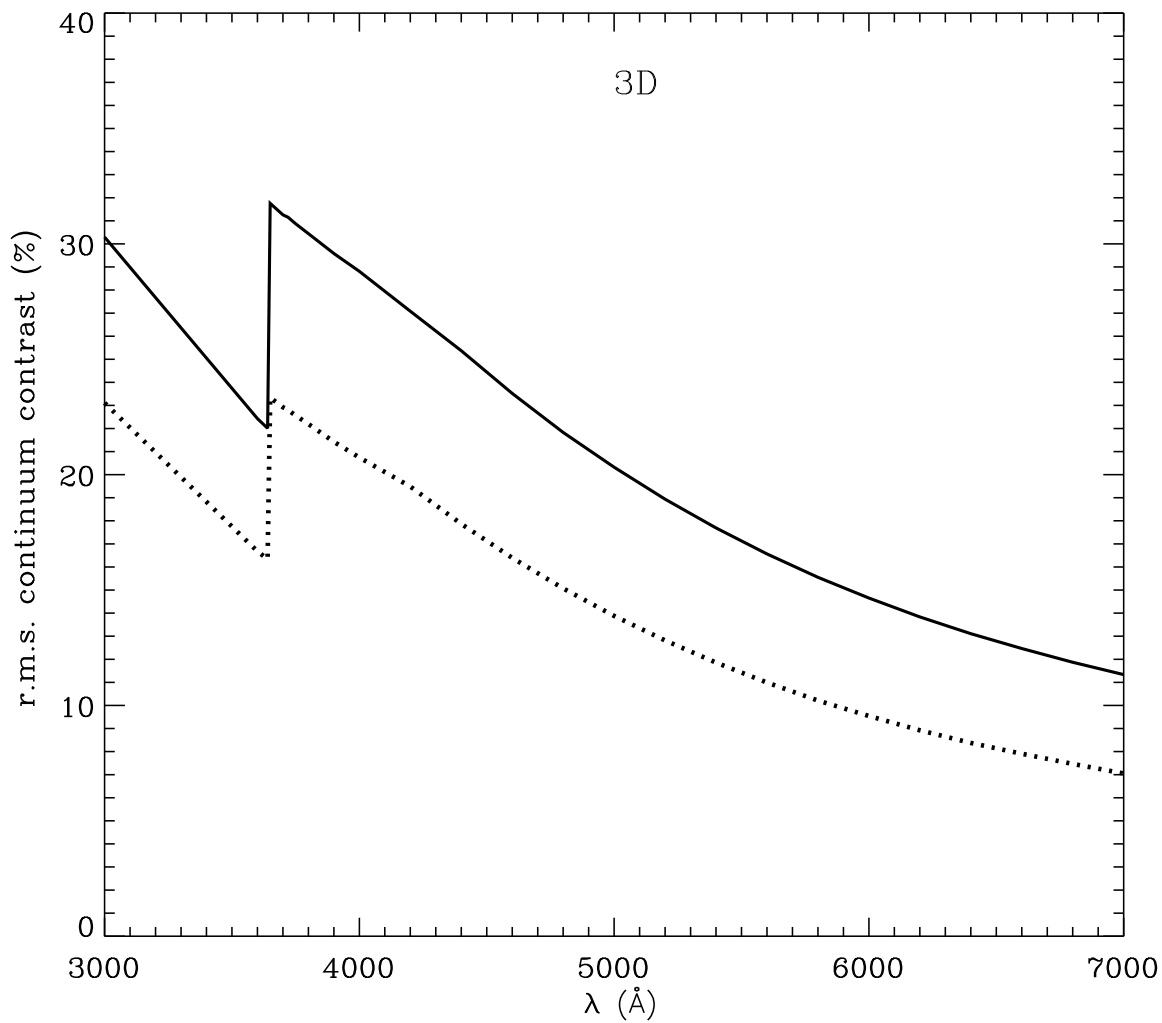}
\caption{The wavelength variation of the RMS contrast of solar granulation calculated in the 3D model.
Solid line: from the raw simulations (i.e., without any smearing). 
Dotted line: after accounting for the degradation produced by the solar space telescope Hinode, without introducing defocusing and assuming that at all wavelengths we have a performance identical to that corresponding to the {\em Hinode} spectropolarimeter (which measures the Stokes parameters of the Fe {\sc i} lines around 6302 \AA). Note that at 6301 \AA\ the calculated RMS contrast after the degradation effect produced by {\em Hinode} is 8.6\%. 
\label{fig:RMS}}
\end{figure}

\clearpage

\begin{figure}
\plottwo{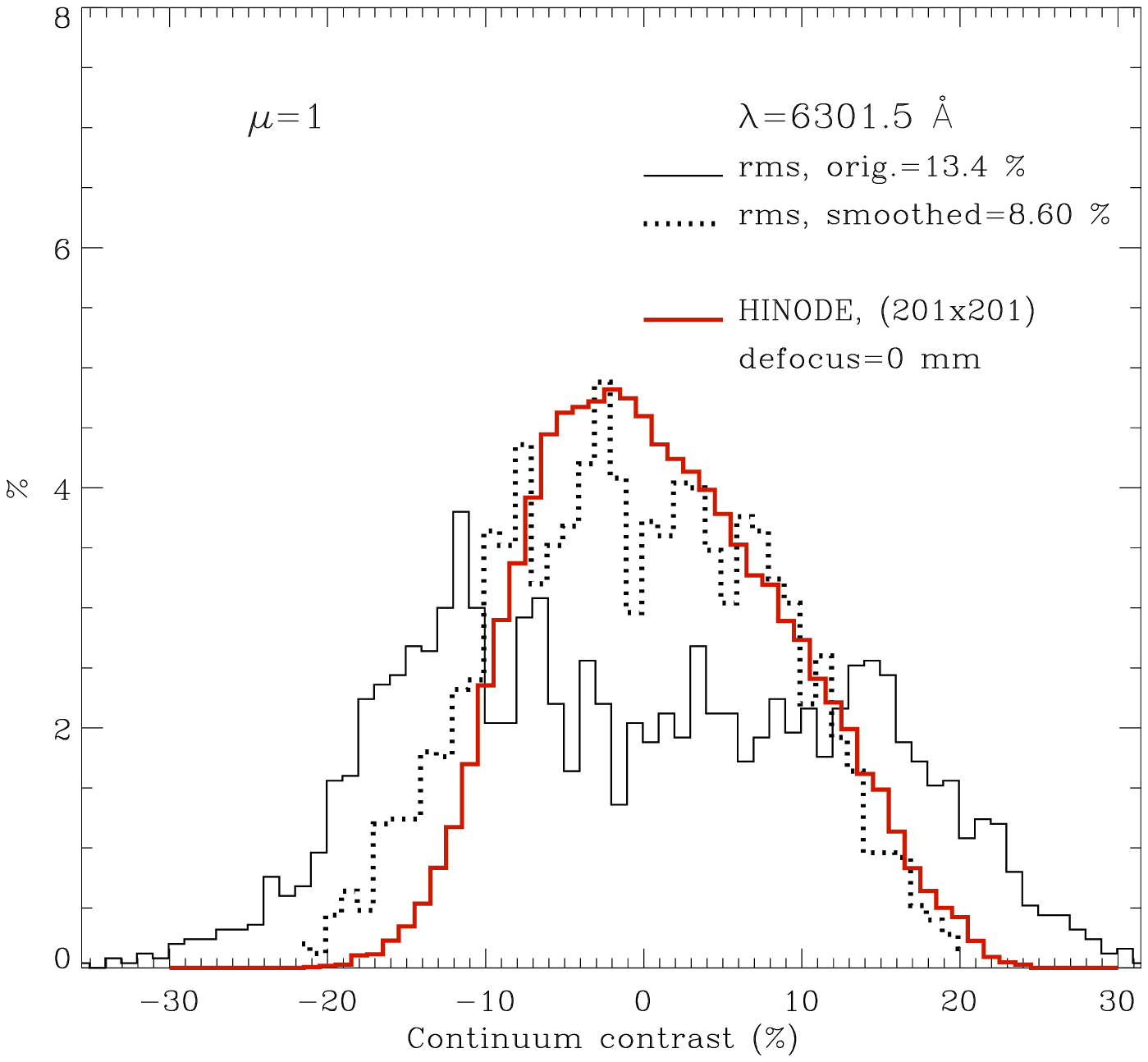}{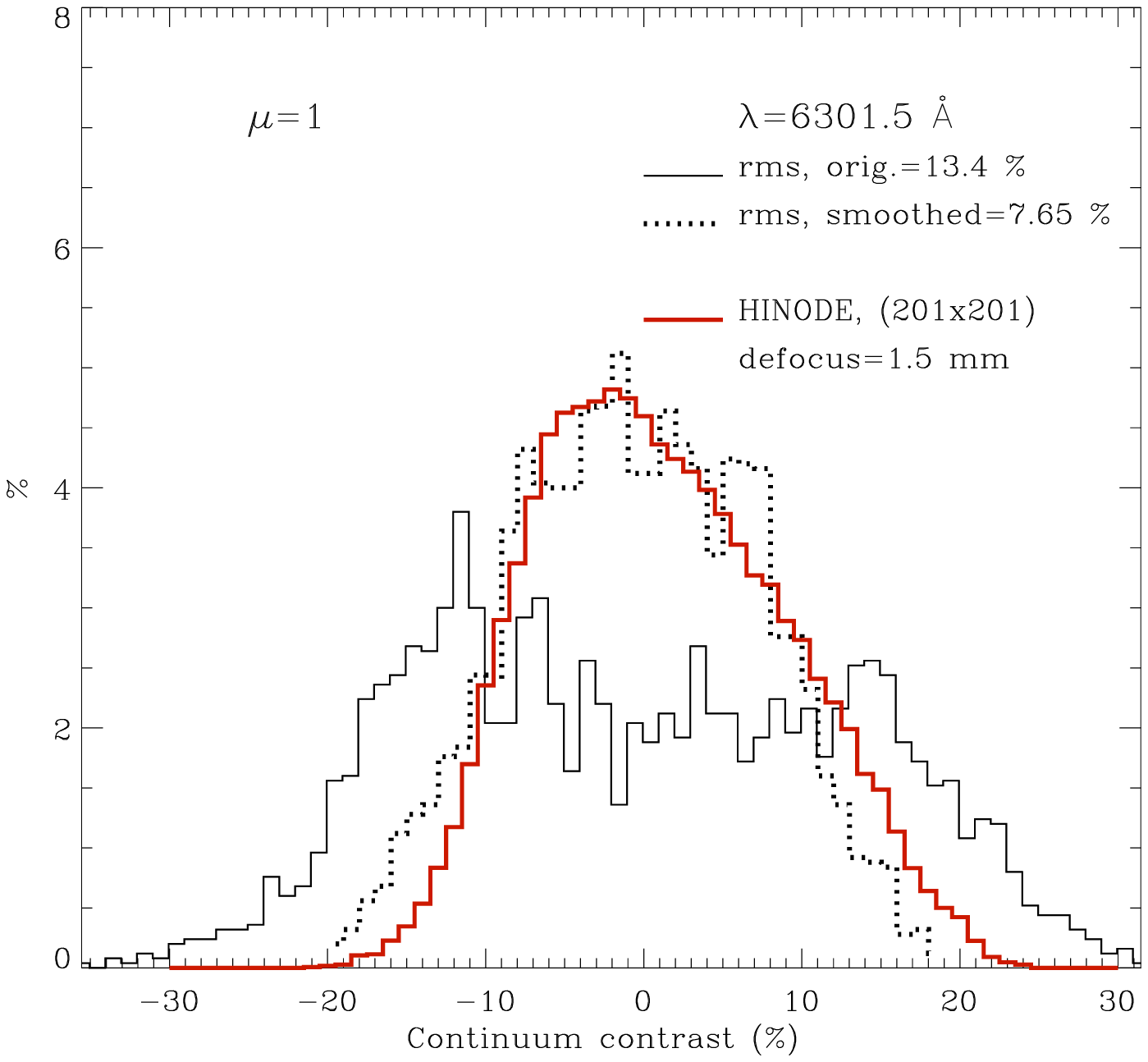}
\caption{Histograms of the emergent radiation in a continuum window close to 6302 \AA\ from the raw simulation (the solid-line with a doubled-peaked distribution), after the degradation introduced by the solar space telescope {\em Hinode} (dotted lines) with the defocusing 
given in the corresponding panel (dotted lines). The thick solid line with a single peak 
has been obtained from a real observation of the ``quiet" Sun taken with the spectropolarimeter 
onboard {\em Hinode} (the solid line with a single peak, which appears in red color in the electronic version of this paper). 
\label{fig:histogram}}
\end{figure}

\clearpage

\begin{figure}
\plotone{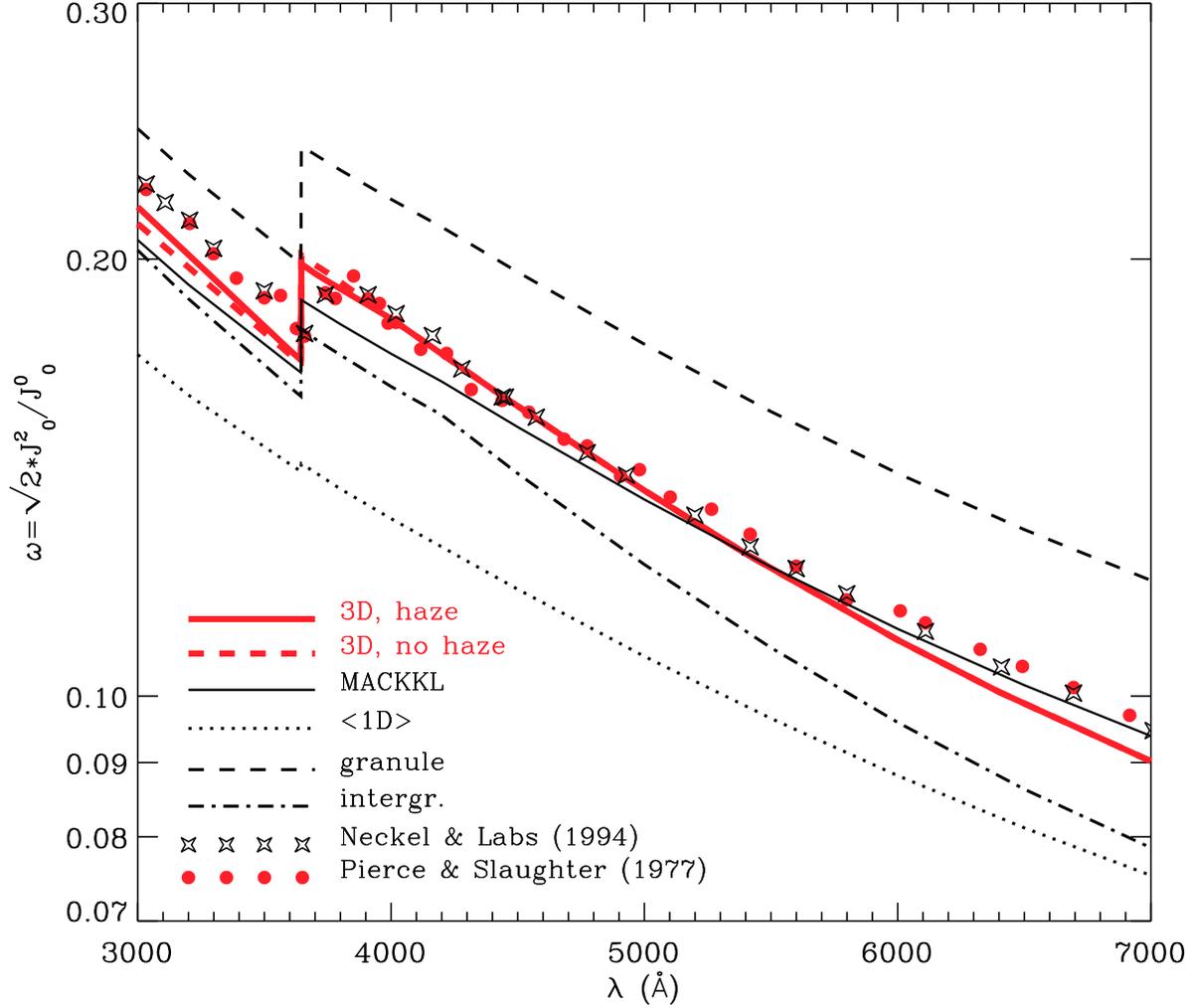}
\caption{The wavelength variation of the anisotropy factor. Symbols: the empirical values obtained from the center-limb observations by Pierce \& Slaughter (1977) and Neckel \& Labs (1994). The various curves give the calculated anisotropy factor at sufficiently large heights
in the indicated models, so as to be sure that the corresponding {\em incoming} continuum radiation is negligible.
\label{fig:anisotropy}}
\end{figure}

\clearpage

\begin{figure}
\plottwo{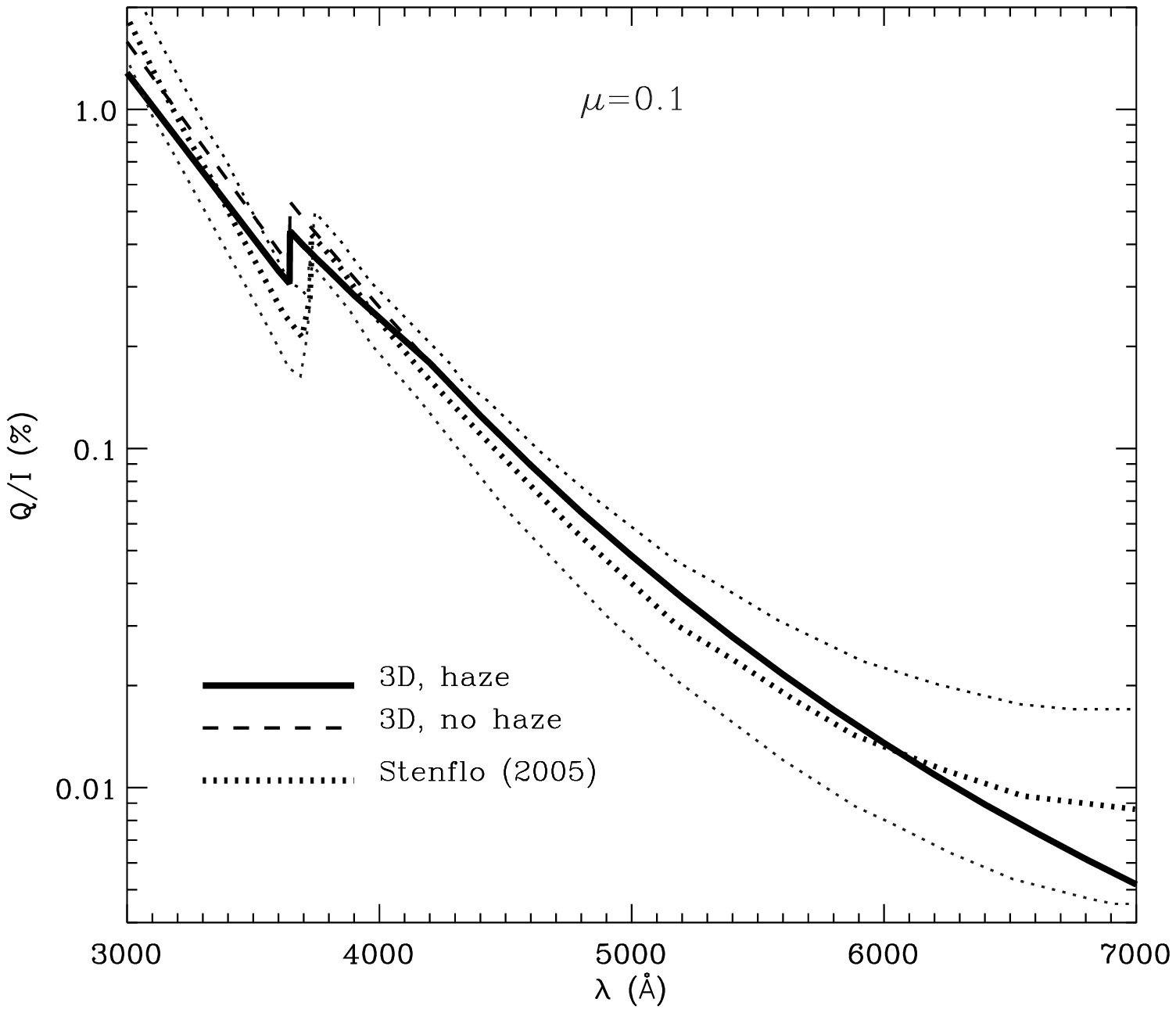}{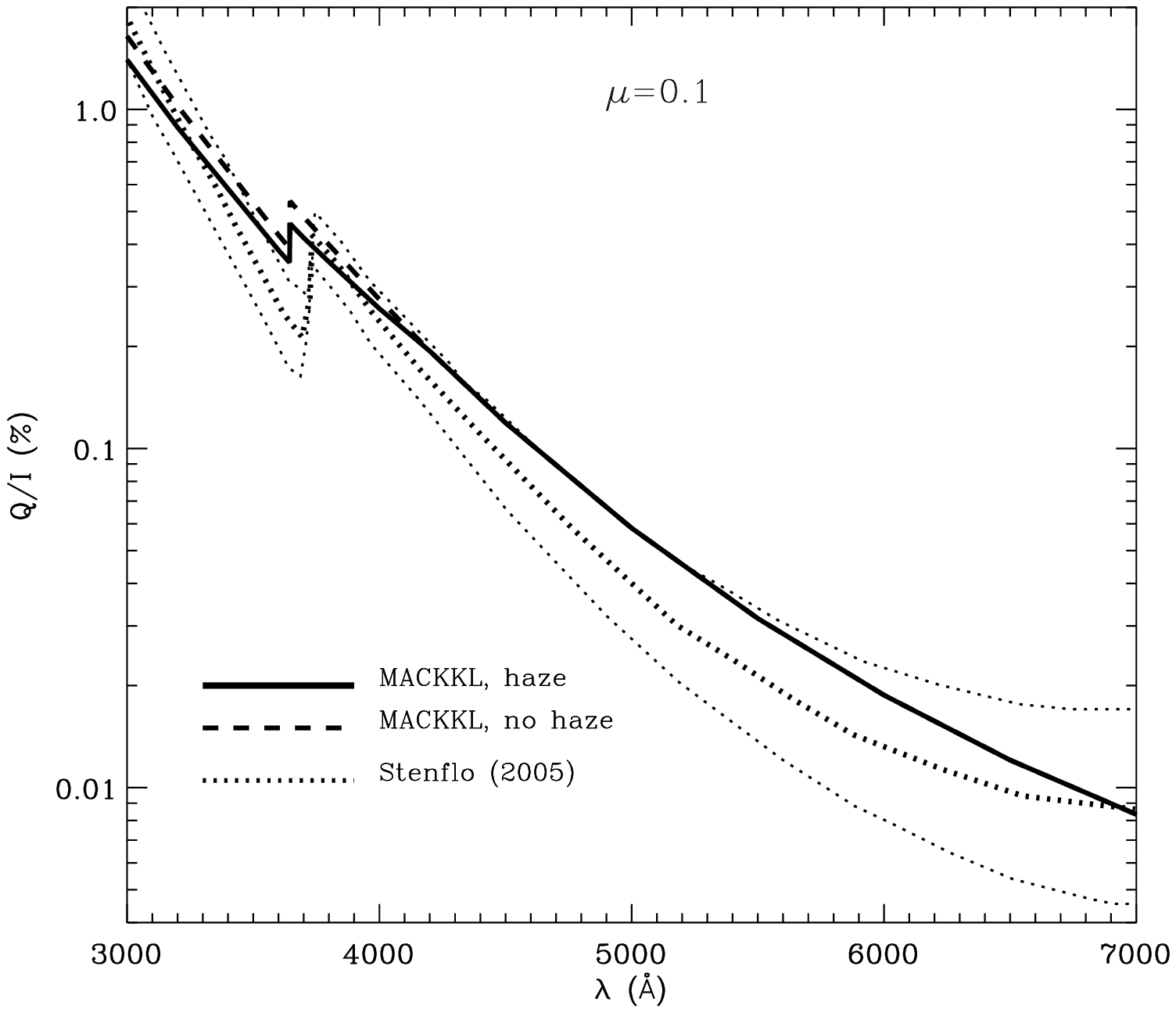}
\caption{The wavelength variation of the polarization of the Sun's continuous spectrum. In both panels the three dotted lines correspond to Stenflo's (2005) semi-empirical determination, with the central curve showing the most likely representation. 
The solid and dashed lines show the results of our radiative transfer calculations in the 3D model (left panel) and in MACKKL model (right panel). 
\label{fig:polcon-1}}
\end{figure}

\clearpage

\begin{figure}
\plotone{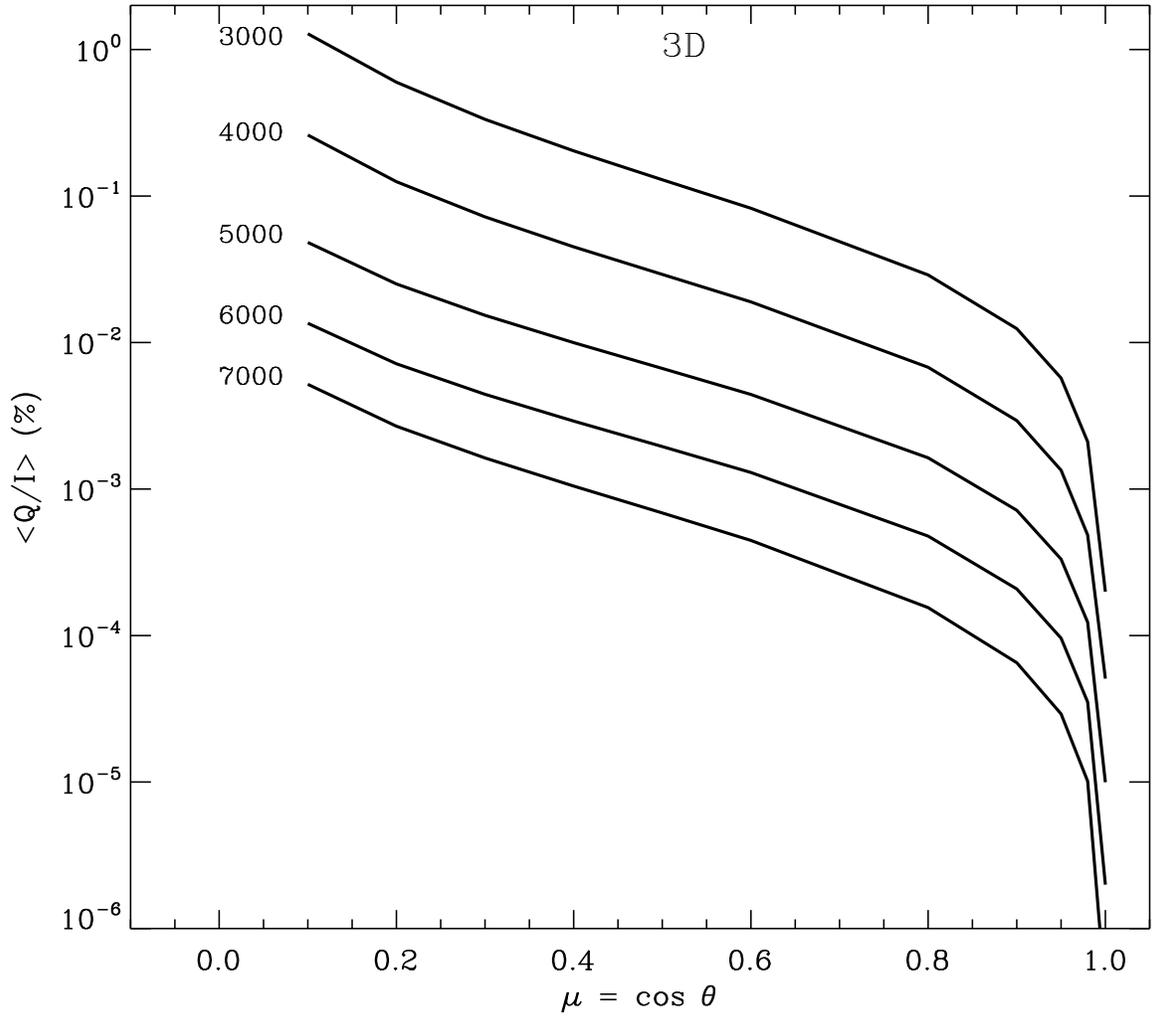}
\caption{Overview of the functional behavior of the continuum polarization calculated in the 3D model. Each curve gives the center-limb variation of the continuum polarization for the indicated wavelengths. 
\label{fig:polcon-3}}
\end{figure}

\clearpage

\begin{figure}
\plotone{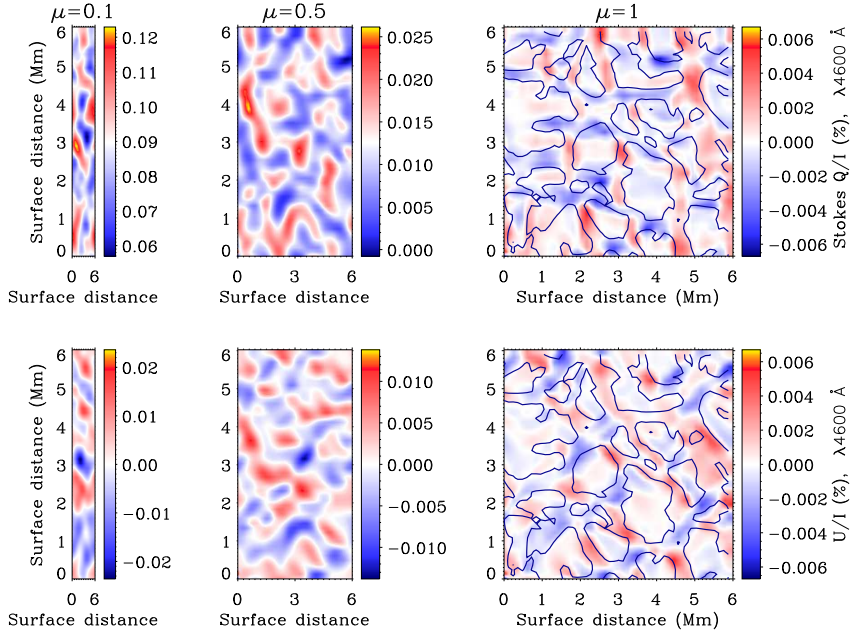}
\caption{The emergent $Q/I$ (top panels) and $U/I$ (bottom panels) at 4600 \AA\ calculated for three line of sights in the 3D model and accounting for the diffraction limit effect of a 1-m telescope. The positive reference direction for Stokes $Q$ lies along the vertical direction of the corresponding panel, which for the $\mu=0.1$ and $\mu=0.5$ cases coincides with the parallel to the limb of the chosen solar atmospheric model. Note that we have taken into account the projection effects by means of which the off-disk-center images appear contracted by a factor $\mu$ along the  horizontal direction of the figure panels. Note also that the ``surface distances" given in the plots measure the true separation between the points on the actual surface of the solar model. The solid-line contours in the $\mu=1$ panels delineate the (visible) upflowing granular regions.\
\label{fig:resolve-1}}
\end{figure}

\clearpage

%%%%%%%%%%%%%%%%%%%%%%%%%%%%%%%%%%%%%%%%%%%%%%%%%%%%%%%%%%%%%%%%%
% The bibliography
%%%%%%%%%%%%%%%%%%%%%%%%%%%%%%%%%%%%%%%%%%%%%%%%%%%%%%%%%%%%%%%%%

\end{document}